\documentclass[a4paper,fleqn,usenatbib]{mnras}
\usepackage{hyperref}
\usepackage[T1]{fontenc}
\usepackage{natbib}
\usepackage{epsfig}
\usepackage{ae,aecompl}
\usepackage{graphicx}    
\usepackage{amsmath}     
\usepackage{array}  
\usepackage{arydshln}   
\usepackage{amssymb}     
\usepackage{gensymb}     
\usepackage{pdflscape}
\usepackage{comment}
\usepackage{subcaption}
\usepackage{breakurl}
\usepackage{arydshln}
\usepackage{array}
\usepackage{booktabs}
\usepackage{newtxtext,newtxmath}
\hypersetup{pdftitle=Spectro-Temporal Study of TXS 1700+685}

\title[Quasi periodic oscillations in ZTF blazars]{Probable low-frequency quasi-periodic oscillations in blazars from the ZTF survey}
\author[Banerjee et al.]{Anuvab Banerjee$^{1}$,
Vibhore Negi$^{2,3}$\thanks{E-mail: vibhore.negi18@gmail.com},
Ravi Joshi$^{4,5}$,
Nagendra Kumar$^{6}$,
Paul J. Wiita$^{7}$,
Hum Chand$^{8}$,
\newauthor
Nikita Rawat$^{2,3}$,
Xue-Bing Wu$^{5,9}$,
Luis C. Ho$^{5,9}$ \\
\\
$^{1}$S.N. Bose National Centre for Basic Sciences, Block JD, Sector 3, Salt Lake, Kolkata, 700106, India\\
$^{2}$Aryabhatta Research Institute of Observational Sciences (ARIES), Manora Peak, Nainital, 263002, India\\
$^{3}$Department of Physics, Deen Dayal Upadhyaya Gorakhpur University, Gorakhpur, 273009, India\\
$^{4}$Indian Institute of Astrophysics, Koramangla, Bangalore, 560034, India\\
$^5$Kavli Institute for Astronomy and Astrophysics, Peking University, Beijing, 100871, China\\
$^{6}$No. 1006, Santosh M'house,  9th Cross, Divanarapalya, Bangalore, 560054, India\\
$^{7}$Department of Physics, The College of New Jersey, 2000 Pennington Rd., Ewing, NJ, 08628-0718, USA \\
$^{8}$Central University of Himachal Pradesh, Dharamshala, 176215, India\\
$^{9}$Department of Astronomy, School of Physics, Peking University, Beijing, 100871, China}
\date{Accepted XXX. Received YYY; in original form ZZZ}
\pubyear{2022}
\begin{document}
\label{firstpage}
\pagerange{\pageref{firstpage}--\pageref{lastpage}}
\maketitle
\begin{abstract}
We investigate the possible presence of quasi-periodic oscillation (QPO) signals in 2103 blazars from the Zwicky Transient Facility (ZTF) time-domain survey. We detect a low-frequency QPO signal in five blazars observed over these 3.8-year-long optical r-band ZTF light curves. These periods range from 144 days to 196 days detected at $\gtrsim
4\sigma$ significance levels in both the Lomb-Scargle periodogram and Weighted Wavelet Z-transform analyses. We find consistent results using the phase dispersion minimization technique. A similar peak is detected in the g-band light curves at a slightly lower significance of 3$\sigma$. Such nearly periodic signals on these timescales in optical wavebands most likely originate from a precessing jet with high Lorentz factor, closely aligned to the observer's line of sight or the movement of plasma blobs along a helical structure in the jet. 
%
%
\end{abstract}


\section{Introduction}
\label{sec:intro}
Blazars are a sub-class of Active Galactic Nuclei (AGNs) which are marked by their powerful parsec-scale relativistic radio jets closely aligned along the observer's line of sight  \citep{1995PASP..107..803U}. Blazars emit copiously across the entire electromagnetic spectrum and their emission dominates the $\gamma$-ray sky \citep{1992ApJ...386..473H}. Blazars are further classified into two sub-categories on the basis of their optical spectra, namely flat-spectrum radio quasars (FSRQs) characterized by the broad emission lines in the optical spectra, and BL Lac objects, marked by essentially featureless continua \citep{1980ARA&A..18..321A}. \par 

The broadband spectral energy distribution (SED) of blazars features the characteristic double-hump configuration in the logarithmic luminosity–frequency plane. The low-energy peak is produced by the synchrotron emission of the relativistic particles contained in the jet, and the second high energy hump is usually considered to be produced by the inverse Comptonization of  lower energy seed photons. In this leptonic scenario, there are two basic possibilities for the  origin of the seed photons. In the synchrotron self-Compton (SSC) scenario the same population of electrons produced by the synchrotron process are inverse Comptonized to produce high-energy emission \citep{1992ApJ...397L...5M}, whereas in the external Comptonization (EC) models, the seed photons originate from various regions 
external to the jets, e.g., accretion disc \citep{dermer1993model}, broad-line region \citep{sikora1994high}, or the surrounding
dusty torus \citep{blazejowski2000comptonization}. The synchrotron peak frequency is a key parameter, on the basis of which the BL Lac sources are sub-divided into low- (LBL; $\log{\nu_{\text{peak}}} < 14$ Hz), intermediate- (IBL; $14 \text{ Hz}\leq \log{\nu_{\text{peak}}} \leq 15$ Hz) and high- (HBL; $\log{\nu_{\text{peak}}} > 15$ Hz) synchrotron peaked blazars \citep{abdo2010spectral}. \par 


Blazars show substantial variability across the entire electromagnetic window, ranging from radio to high energy $\gamma$-rays, which is strongly enhanced by the relativistically beamed jet. In general, such variability is stochastic or aperiodic in nature, but in recent times several instances of quasi-periodicities in different electromagnetic wavebands have been reported \citep[e.g.][]{lachowicz20094,rani2009nearly,bhatta2016detection,smith2018evidence}.
%
%
%
Such quasi-periodic oscillation (QPO) features in blazars have been claimed to be observed on diverse time-scales in optical, radio, X-ray, and $\gamma$-ray wavebands. Several early nominal detections of QPO features, such as a $\sim$15-min periodicity in 37 GHz radio monitoring \citep{valtaoja198515}, a $\sim$23-min periodicity observed in the optical band \citep{carrasco1985periodicity}, and a $\sim$1 day periodicity of S5 0716+714 during coordinated radio and optical campaign \citep{quirrenbach1991correlated}, were however, plagued by improper modeling of the underlying broadband noise \citep[see][]{vaughanUttley2006detecting}. The first more robust detection of a QPO feature used a 91 ks observation of the Narrow Line Seyfert 1 galaxy RE J1034+396 by \textit{XMM-Newton} was performed by \citet{gierlinski2008periodicity}, where they reported a $5.6\sigma$ detection of a $\sim$1 hr QPO. Subsequently, nominally strong detections of several other QPOs in different AGNs using X-ray monitoring data have been reported: $\sim$7$\sigma$ detection of a 4.6 hr QPO in PKS 2155$-$304 \citep{lachowicz20094}, a $> 3\sigma$ detection of a 3.8 ks QPO in 1H 0707$-$495 \citep{pan2016detection}, a $\sim$1 hr QPO of >3$\sigma$ detection in MCG$-$06-30-15 \citep{gupta2018possible}, and a 3.3$\sigma$ detection of a $\sim$2 hr QPO in the active galaxy MS 2254.9$-$3712 \citep{alston2015discovery}. However, since most such detections lasted for only a few cycles, their statistical significance was probably overestimated \citep{covino2019gamma}. \par 

 A remarkable correlation is seen between QPO frequency and the mass of the central object across several orders of magnitude, ranging from stellar mass black holes in binary systems to supermassive black holes in AGNs \citep[e.g.][]{zhou2014universal}. Such a correlation strongly suggests that the underlying accretion flow properties are similar in these two classes of sources, and a $\frac{1}{M}$ scaling relation is expected if the oscillatory features are connected to the characteristic length scales of the system \citep{smith2018evidence}. However, this would indicate that the low-frequency QPO ($\sim $1 Hz QPO in the case of a $10 M_\odot$ low mass X-ray binary) would occur at time-scales of months to years in AGN \citep{Ackermann_2015,zhang2017gamma,bhatta2019blazar}. Detecting these QPOs would invariably require long duration, high cadence, and monitoring of the sources. With the advent of \textit{Fermi}-LAT, which enables the monitoring of the $\gamma$-ray sky as frequently as every 3 hr for the brightest blazars, as well as high cadence optical monitoring programs, it is possible to make significant progress in the exploration of such low-frequency, or long time-scale, QPOs. Examination of $\gamma$-ray and optical/NIR lightcurves of six blazars from Rapid Eye Mounting telescope photometry yielded a statistically significant peak at a year-like time-scale in PKS 2155$-$304 \citep{sandrinelli2016quasi}. The detection of optical QPOs, particularly on longer time-scales, has always been challenging, owing to the normally sparse sampling by  ground-based monitoring instruments and data gaps induced by poor weather and the daytime transit of most AGNs. Enabled by the high precision and nearly continuous monitoring provided by the \textit{Kepler} exoplanet satellite, \citet{smith2018evidence} found a significant detection of a $\sim$44 days QPO in KIC 9650712, a narrow-line Seyfert 1 galaxy, which was confirmed by \citet{phillipson2020complex} who employed an independent approach to analysing those \textit{Kepler}  data. \par

A number of physical models have been invoked to explain the observed AGN QPO features, including bending or twisting of the relativistic jets \citep[e.g.][]{camenzind1992lighthouse,raiteri2017blazar}, the precession of the jet due to interaction in a supermassive black hole binary (SMBH) system \citep[e.g.][]{graham2015systematic}, warped accretion disc structure in a  binary SMBH system \citep[e.g.][]{ulubay2009self}, the movement of plasma blobs along a helical structure in the jet \citep[e.g.][]{sarkar2021multiwaveband}, and kink instabilities in magnetized jets \citep[e.g.][]{2022Natur.609..265J}. The multiplicity of plausible  explanations calls for both more QPO detections and a detailed investigation of their multi-wavelength variability properties in order to put real constraints on the plausible origin involving physical mechanisms within the jet and/or accretion disc, as well as their possible coupling. \par 

Here, we report the probable detections of elusive lower frequency QPO signatures in a few  blazars over the rather long time baseline provided by the Zwicky Transient Facility (ZTF) r-band   observations which span nearly four years with measurements made every few days.
Optical observations over such time spans could reveal several cycles of such putative QPO variability and so are very well suited  to explore the presence, strength, and evolution of the low-frequency quasi-periodic variations. This paper is organized as follows. The data set is discussed in Section 2, followed by the results in Section 3. We
describe  various possible scenarios to explain the QPO signatures in Section 4. Finally, a summary is given in Section 5. Throughout
this paper, we adopt a set of cosmological parameters as follows: $\rm H_0 = 70\ km \ s^{-1} \ Mpc^{-1}, ~\Omega_m = 0.30, ~\Omega_{\Lambda} = 0.70$.


\begin{table}
	\centering
	\caption{Best fit parameter values}
	\label{tab:modelpars}
	\begin{tabular}{ccl} 
		\hline
		Source & Best-fit & Best-fit\\
             & model & parameters\\
		\hline
             & & $A = 39.21$\\
              & & $\nu_b = 0.012$\\
		J092915+501336 &  BPL & $\alpha_1 = -1.92$\\
            & & $\alpha_2 = 2.53$\\
             & & $c = 1.65\times10^{-5}$\\
             \hline
             & & $A = 0.51$\\
              & & $\nu_b = 0.021$\\
		J092331+412527 &  BPL & $\alpha_1 = -0.92$\\
            & & $\alpha_2 = 2.33$\\
             & & $c = 9.61\times10^{-5}$\\
		J101950+632001 & PL & $A = 1.71\times10^{-5}$\\
            & & $\alpha = 1.41$\\
            J173927+495503 & PL & $A = 1.01\times10^{-5}$\\
            & & $\alpha = 1.73$\\
            J223812+274952 & PL & $A = 3.00\times10^{-5}$\\
            & & $\alpha = 1.33$\\
		\hline
	\end{tabular}
\end{table}
     
\section{Data and Sample Selection}
Our preliminary sample of blazars was taken from the Roma-BZCAT catalog \citep[]{2015Ap&SS.357...75M}, which consists of a total of 3561 blazars. We searched for the r-band light curves for all these 3561 blazars in the  ZTF (\citealt{2019PASP..131a8003M}) time-domain survey, the 10th ZTF public data release, and found 2751 of these blazars to be present in this database. The ZTF uses the 48-inch Samuel Oschin Schmidt telescope with a field of view of 48 $deg^2$ to map the sky in g, r, and i optical bands with a typical exposure time of 30 seconds,
reaching a magnitude limit of $\sim$20.5 in r-band (\citealt{2019PASP..131a8002B}). The average cadence of the survey is about three days over a period of 3-4 years, so ZTF is extremely well suited for studying AGN variability \citep[see][]{2022MNRAS.510.1791N}.
Note that ZTF
assigns a unique observation ID to a source observed in a particular field, filter, and CCD-quadrant independently.
Here, to avoid any spurious  variability based on varying calibrations on different CCD-quadrants (see, \citealt{2021AJ....161..267V}), we only examined the light
curve corresponding to the observation ID with the maximum
number of data points. To only include data obtained in good observing conditions, we further applied the quality score of
catflags score = 0, listed in ZTF documentation. 
For the sources having multiple intranight observations, we averaged all the data points observed on same day and used the daily averaged light curves for further analysis. \par

Further, to avoid large data gaps and to have a sufficiently large number of  data points to robustly detect any periodic signature, we included only the sources with at least 100 data points during the entire set of observations. This leaves us with a sample of 2103 blazars observed by ZTF with light curves good enough to search for QPOs. In the next section, we search for any kind of quasi-periodic behaviour in the r-band light curves of these 2103 sources. \\

\begin{table*}
	\caption{Probable cases of QPO in blazars from the ZTF survey.}
	\label{table:results_table_qpo}
	\centering
	\begin{tabular}{c c c c c c  c c c}
	\hline\hline
No	&	Source & RA & Dec.  & Peak frequency$^\dagger$    & Period (LSP) &   Period (WWZ) &   Period (PDM) & Global Sig.\\
(\#)	&	         &(deg.) & (deg.) & ($\text{days}^{-1}$)   &    (days)   &   (days) & (days) & {\bf (\%)} \\		
	\hline \\
1	&	J092915+501336 & 142.31429 & 50.22667 & 0.00509 &   196.3$\pm$6.8   &  192.31 & 196.08$\pm$3.85 & 99.4 \\
2	&	J092331+412527 & 140.88042   & 41.42428 & 0.00509  &   196.1$\pm$6.8   &   196.08 & 200.00$\pm$4.00 & 99.7 \\
3	&	J101950+632001 & 154.96196 & 63.33378 & 0.00615 &   162.7$\pm$4.6   &   163.93 & 163.93$\pm$2.69 & 99.8 \\
4	&	J173927+495503 & 264.86408 & 49.91761 & 0.00694  &  144.1$\pm$3.9   &  142.86 & 142.86$\pm$2.04 & 99.3 \\
5	&	J223812+274952 & 339.55358 & 27.83133 & 0.00647 &   154.5$\pm$4.7   &   153.84 & 156.25$\pm$2.44 & 99.4\\\\
	\hline
    \multicolumn{8}{l}{}\\
    \multicolumn{8}{l}{}
	\end{tabular}
\end{table*}

\section{Analysis and Results}

Here, we employ three different mathematical techniques in order
to detect and quantify any statistically significant oscillation.


\subsection{Weighted Wavelet Z-transform}

The wavelet transform method has been widely employed in  analyses of blazar time-series \citep[e.g.][]{lachowicz20094,mohan2015kinematics,bhatta201372,bhatta2016detection}. This approach determines any periodicities by fitting to sinusoids; however, it also offers the ability to  localize the waves in both time and frequency space to explore possibly transient QPOs  \citep[e.g.][]{bravo2014wavelets}. By examining any evolution of the signal's frequency and amplitude it is a powerful tool for investigating whether such oscillations gradually develop, evolve, and dissipate over time.\par

 In brief, the Weighted Wavelet Z-transform (WWZ) method convolves a light curve with a time- and frequency-dependent kernel and decomposes the data into time and frequency domains to create a WWZ map. We use the Morlet kernel \citep{grossmann1984decomposition}  which has the functional form
\begin{equation}
    f[\omega(t-\tau)] = \exp[i\omega(t-\tau) - c\omega^2(t-\tau)^2],
\end{equation}
corresponding to which the WWZ map is
\begin{equation}
    W[\omega,\tau;x(t)] = \omega^{1/2}\int{x(t)f^{*}[\omega(t-\tau)]dt},
\end{equation}
where $f^{*}$ is the complex conjugate of the Morlet kernel $f$, and $\omega$ and $\tau$ are respectively the frequency and the time-shift. This kernel acts as a windowed
discrete Fourier transform which contains a frequency-dependent window of
size $\exp{(-c\omega^2(t-\tau)^2)}$. The WWZ map has the advantage of
being able to detect statistically significant periodicities, as well as the time spans of their persistence. \par 

For the purpose of our analysis, we have considered only those cases where a strong concentration of power within a narrow window of frequency is detected, and the peaked components pertaining to such cases are successively tested for their statistical significance, as will be discussed in the subsequent sections. Most of the wavelets show the most significant concentration of power toward the lower frequency region, i.e., timescales of more than $\sim$100 days. We note that the time-series are associated with gaps of at least a few tens of days duration, hence we have considered only those cases where a concentration of power is observed throughout the domain of observation. A total of 169 such cases were identified, and a rigorous estimation of their significance was done as explained in the next section. 
 
\subsection{Lomb-Scargle periodogram and testing the peak significance}
 The traditional Lomb-Scargle periodogram (LSP) method is one of the most efficient and widely used methods for periodicity search \citep{1976Ap&SS..39..447L,scargle1982studies}. Though it is a variant of the standard discrete Fourier transform (DFT) method, the advantage of the LSP is that it attempts to account for the data gaps and irregularities by the least-square fitting of the sinusoidal waves of the form $X(t) = A\cos{\omega{t}} + B\sin{\omega{t}}$ of the data by $\chi^2$ statistics, which reduces the effect of the noise on the signal and also provides a measure of the significance of detected periodicity  \citep[e.g.][]{zhang2017revisiting,zhang2017possible}. \par 
 
Blazars typically exhibit a red-noise type variability feature in temporal frequency space, such that the periodogram can be represented by a power spectral density (PSD) with the functional form of a power-
law (PL) as $P(\nu) = A\nu^{-\alpha}$ where $\nu$ represents the temporal frequency, $\alpha > 0$ is the spectral slope.  Alternatively, a bending power law (BPL) might be a better fit, with 
\begin{equation}
    P(\nu) = A\frac{\nu^{-\alpha_1}}{1 + (\nu/\nu_{\text{bend}})^(\alpha_2 - \alpha_1)} + c, 
\end{equation}
 where $\nu_{\text{bend}}$ is the bending frequency and $\alpha_1$ and $\alpha_2$ are the power spectral slopes before and after the bending frequency respectively   \citep[e.g.][]{timmer1995generating,max2014method}. 
 
 Such power-law profiles indicate that high amplitude features in the PSD seen at longer time scales (low-frequency domain) can mimic an actual QPO signal \citep{vaughan2016false}. In order to disentangle the effect of red noise, rigorous estimation of the periodogram peaks must be undertaken before concluding a peaked feature as a true QPO. In our present work, this issue has been addressed by performing Monte-Carlo simulations of the light curves, such that the simulated light curves exhibit the same PSD and flux distribution (PDF) as those of the original light curve \citep{emmanoulopoulos2013generating}. Since the underlying red-noise PSDs of blazar light curves are well approximated by a PL or BPL profile \citep{vaughan2005simple}, the PSD of the original light curve has been approximated using these functional forms, and the final model has been chosen on the basis of the lower $\chi^2$ value. The best-fit PSD models and model parameters are provided in Table \ref{tab:modelpars}. Subsequently, a total of 1000 light curves with the same PSD and PDF as that of the original light curve were simulated using the \texttt{DELightcurveSimulation}\footnote{\url{https://github.com/samconnolly/DELightcurveSimulation}} code \citep{emmanoulopoulos2013generating}. 

\begin{figure*}
\centering
    \includegraphics[width=17cm,height=5cm]{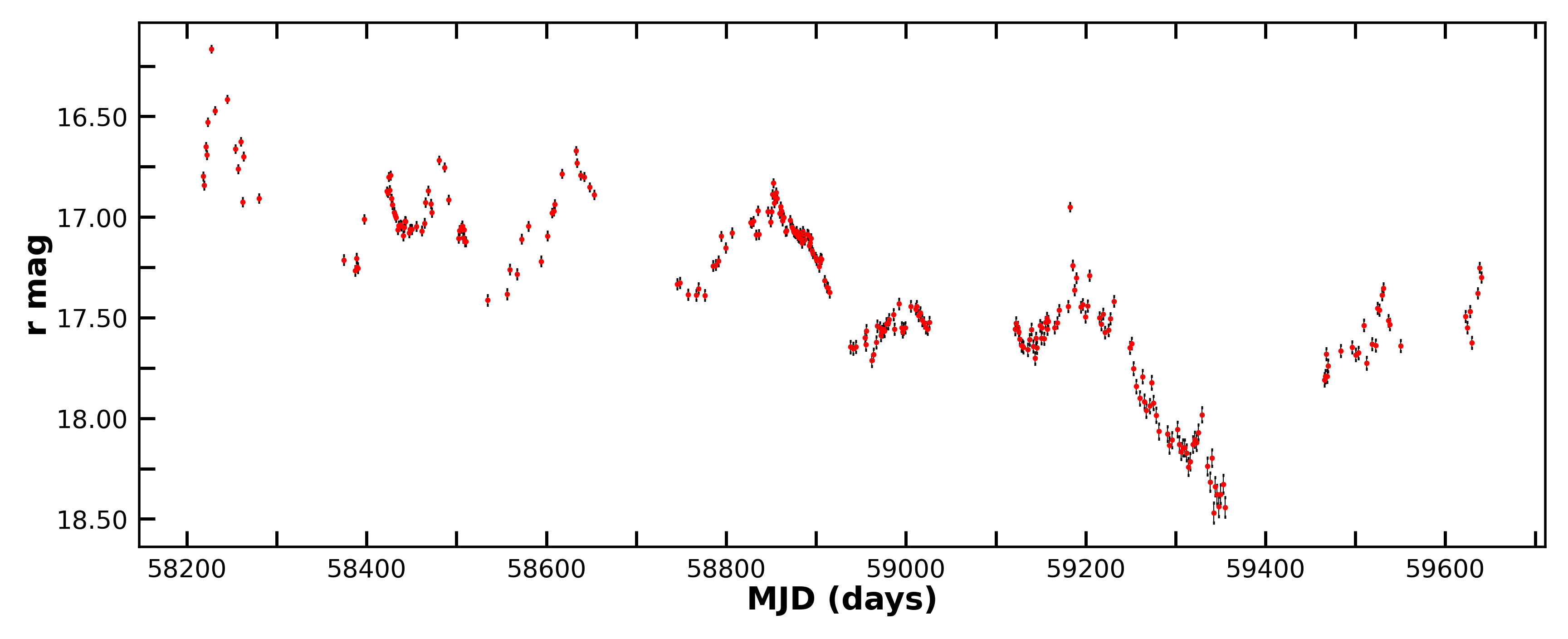}
    \includegraphics[width=17cm,height=10cm]{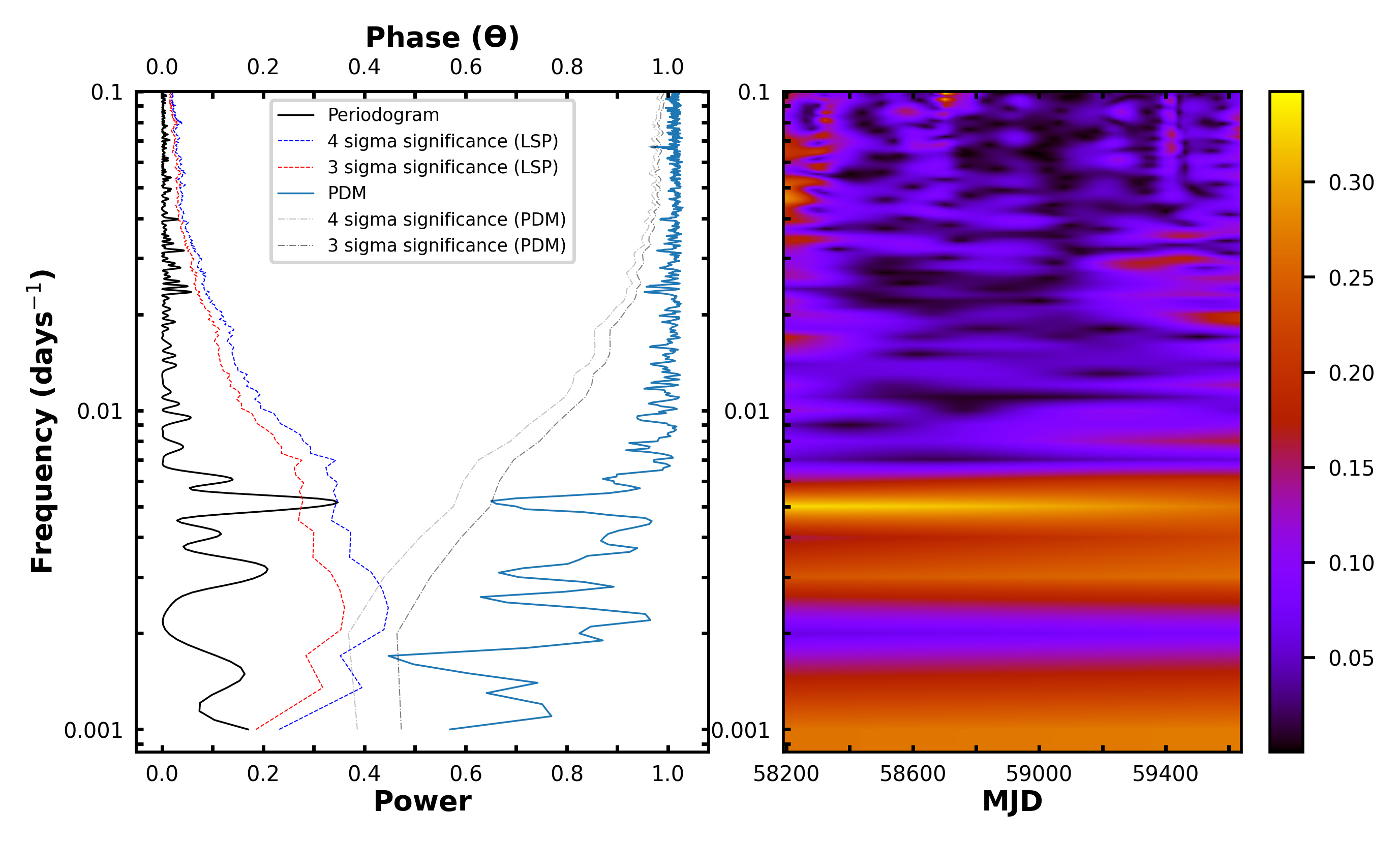}
    \caption{\textit{Upper Panel:} r-band ZTF light curve of blazar J092915+501336 over a time span of 3.8 years. \textit{Bottom left Panel:}  A low-frequency peak in the Lomb-Scargle periodogram at $> 4 \sigma$ statistical significance is detected in the r-band light curve, suggesting the presence of a QPO at 0.0051 days$^{-1}$, and this is supported by the dip below 0.7 in the PDM phase at that frequency, which is $\sim3\sigma$ significance. \textit{Bottom right Panel:} A weighted wavelet Z-transform of this light curve showing a strong power concentration at the same frequency over the entire observing time window.}
    \label{fig:J0929}
\end{figure*}

Our wavelet analysis suggested the plausibility of the presence of low-frequency oscillatory features in a small fraction of the blazars observed by ZTF. We consider the peaks at frequencies above 0.005 $\text{days}^{-1}$ (i.e., <200 days) with $\gtrsim$4$\sigma$ significance level to be the legitimate cases of statistically significant variability features. The frequency bound is selected on the basis of requiring that during the full window of observations, there should be at least 4--5 full cycles of oscillation to identify significant oscillatory features. Variabilities corresponding to longer time scales will not satisfy the criteria for these observations. We also note that the ZTF data have multiple gaps in the light curves so that requiring at least 4-5 cycles helps to pick the genuine QPO candidates. Further, to account for any fake periods arising from gaps in the light curve data, we computed the spectral window power spectrum of the sources using STARLINK PERIOD\footnote{\url{https://starlink.eao.hawaii.edu/devdocs/sun167.htx/sun167.html}} software. The fake periods were estimated by replacing all data points by unity and then computing the Fourier transform of that time series data. We made sure that no periodicity due to the fake gaps fell within 1$\sigma$ of our estimated periods. Only 5 sources satisfy all these criteria, and such cases with statistically significant LSP peaks are listed in Table~\ref{table:results_table_qpo}. \par 
We also checked the false alarm probability (FAP) corresponding to these peaks, which estimates the possibility of getting a peak with the obtained amplitude or higher arising out of the random fluctuations in the data instead of some inherent periodicity. In the absence of any apriori information regarding the periodicity within a particular frequency interval, we checked the `global significance level' of the peaks, also known as the `look-elsewhere effect' or `multiple comparison problem' in statistics \citep{bell2011MNRAS.411..402B}. We calculated the fraction of simulated light curves that show a higher statistical significance of peak detection at any frequency interval, and that provides us the global confidence level of the peaks. In all of our cases, we find that the global confidence level of the detected periodicities are $> 99\%$, and the exact confidence levels are mentioned in Table \ref{table:results_table_qpo}. This lends further credence to the detection of periodic features in these ZTF light curves.


We also searched for the presence of any possible peaked components in the high frequency (i.e., short time interval) zone as well, but the periodograms become noisier in this frequency domain and the power peaks turn out to be at least one order of magnitude smaller than the low-frequency peaks.  Further, given the sampling cadence, a detection of a peaked component corresponding to period of oscillation $\lesssim 20$ days could not be made with high statistical confidence. We therefore do not further consider the presence of any high frequency QPOs.

\begin{figure*}
\centering
    \includegraphics[width=17cm,height=5cm]{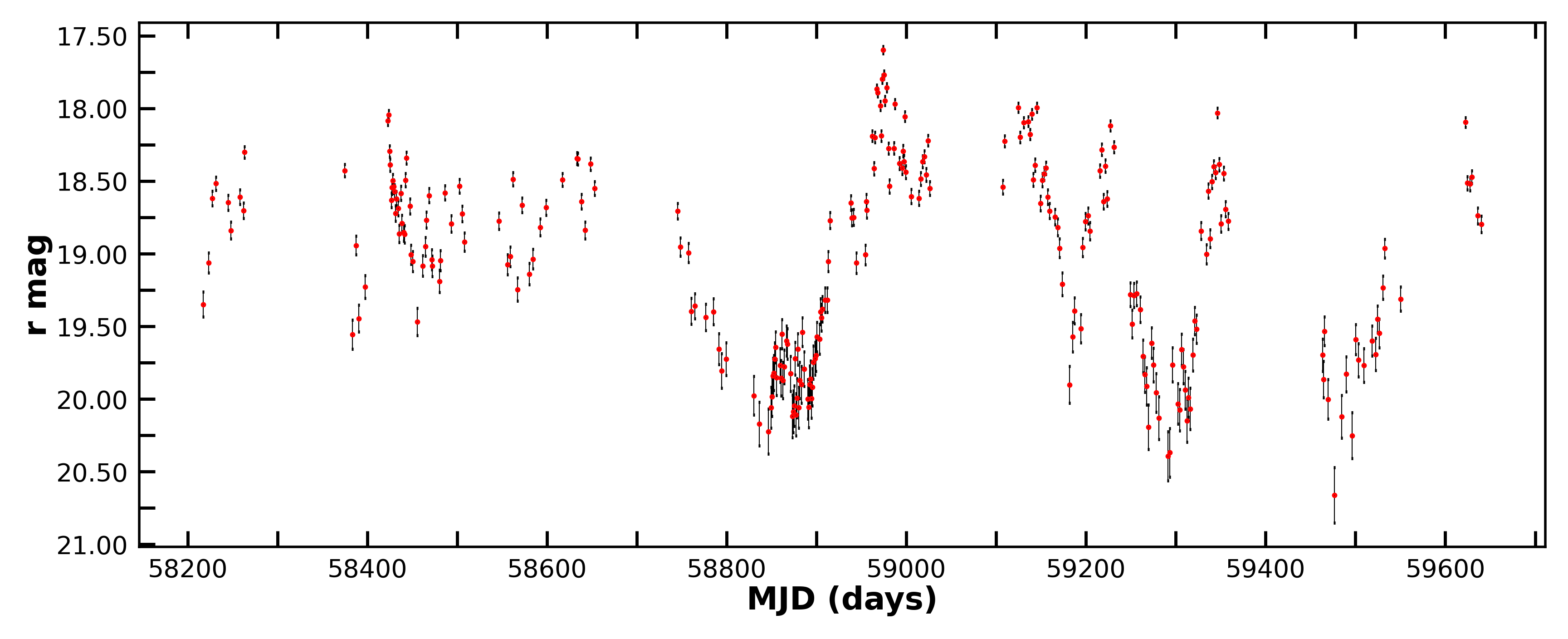}
    \includegraphics[width=17cm,height=10cm]{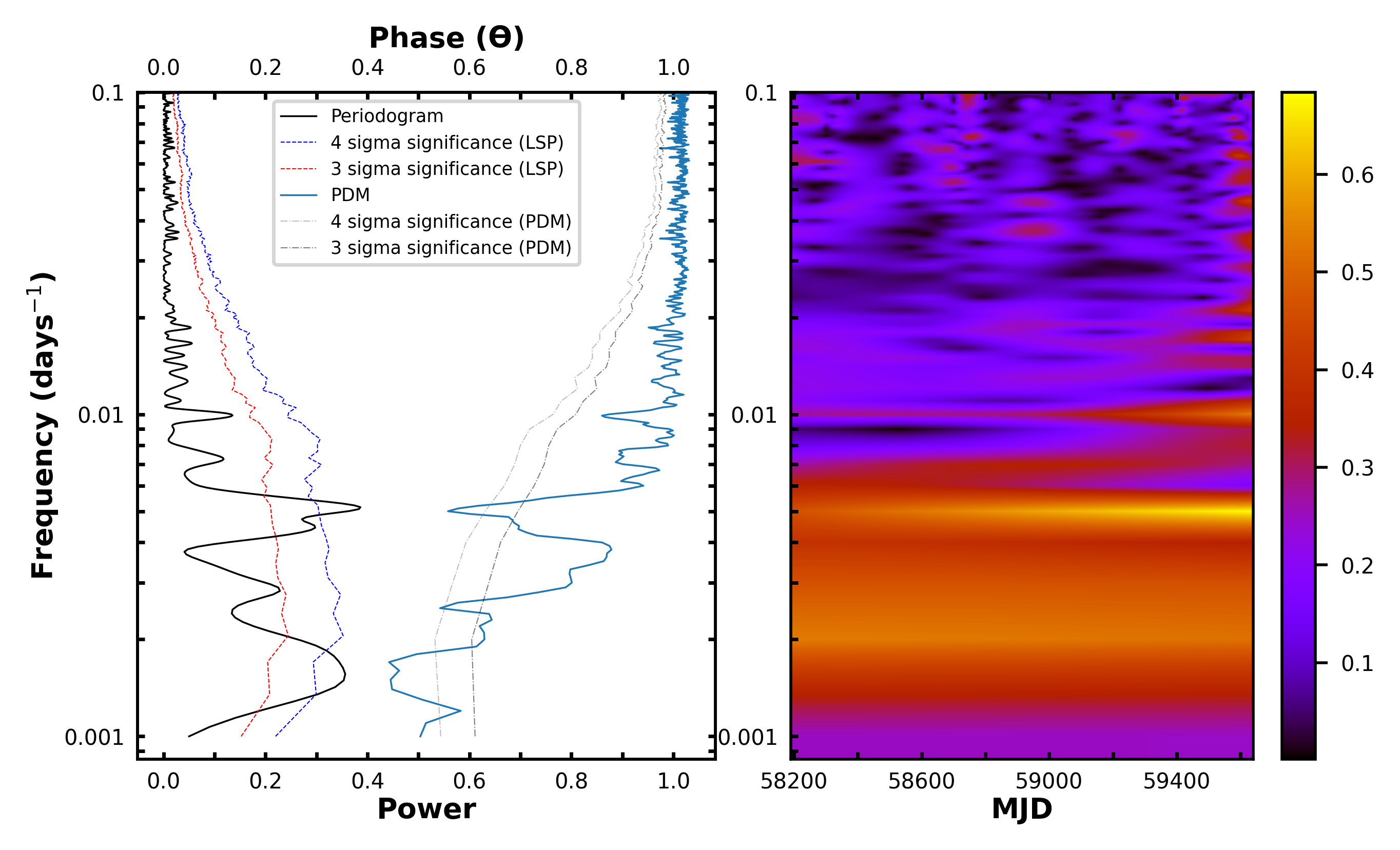}
    \caption{\textit{Upper Panel:} As in Figure~\ref{fig:J0929} for the blazar J092331+412527.  A low-frequency peak in the Lomb-Scargle periodogram at $> 4 \sigma$ statistical significance detected in the r-band light curve, suggesting the presence of a QPO at 0.0051 days$^{-1}$. At this frequency, we observe a $>4\sigma$ significant prominent dip in the phase in our PDM analysis as well. This further strengthens the claim for the presence of a QPO. \textit{Bottom right Panel:} A weighted wavelet Z-transform of this light curve showing a strong power concentration at the same frequency over the entire observing time window.}
    \label{fig:J0923}
\end{figure*}
\begin{figure*}
\centering
     \includegraphics[width=17cm,height=5cm]{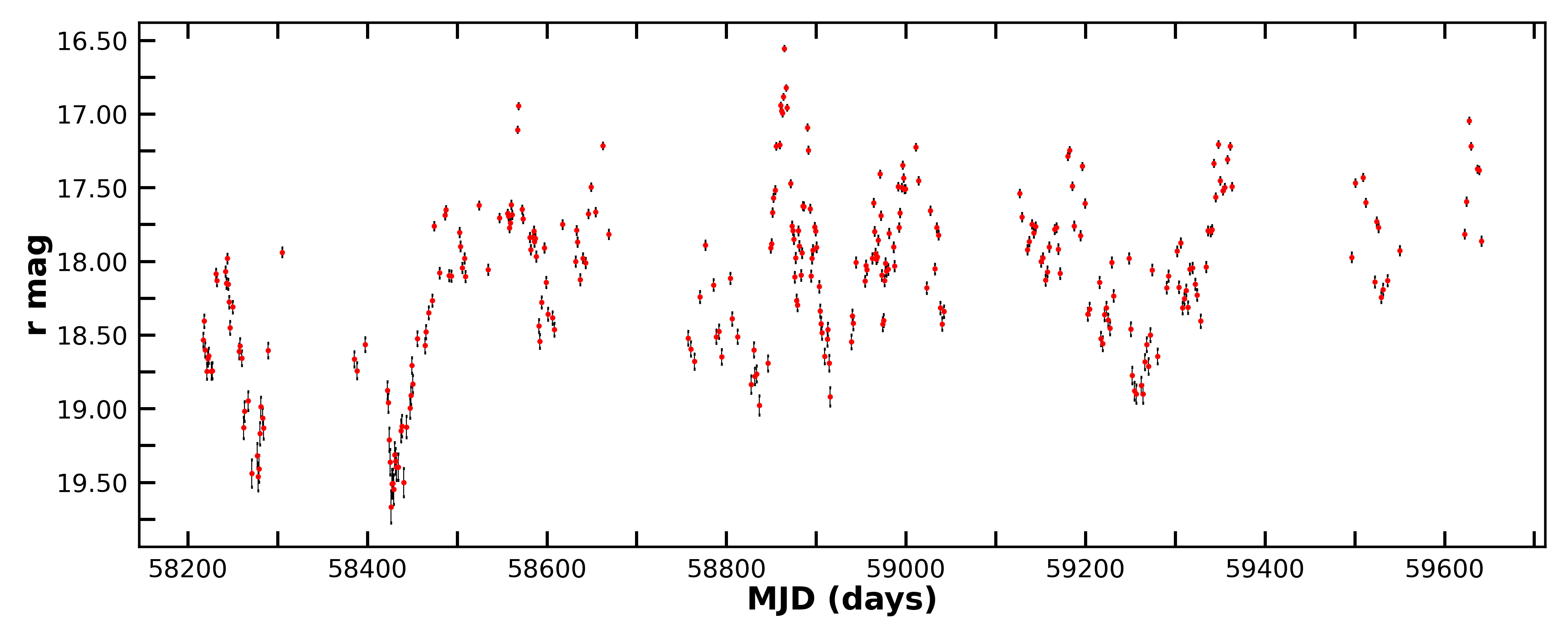}
    \includegraphics[width=17cm,height=10cm]{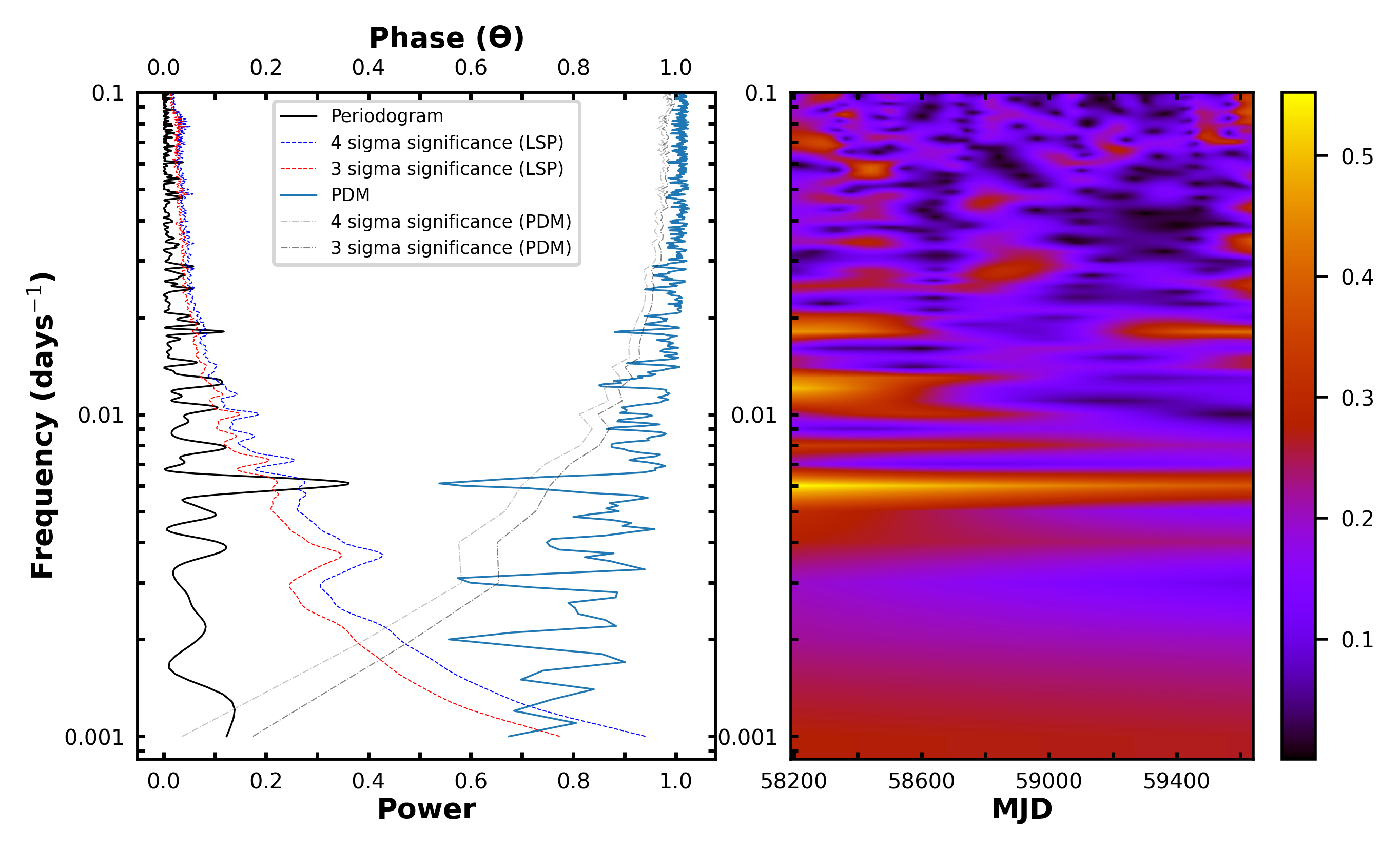}
    \caption{As in Figure~\ref{fig:J0929} for the blazar J101950+632001. A low-frequency peak in the Lomb-Scargle periodogram at $> 4 \sigma$ statistical significance is detected in the r-band light curve, suggesting the presence of a QPO at 0.0062 days$^{-1}$. A \textbf{$>4\sigma$ significant }sharp dip in phase going below 0.6 is also observed from our PDM analysis at this frequency. A strong power concentration at the same frequency is seen in the WWZ plot over the entire observing time window.}
    \label{fig:J1019}
\end{figure*}
\begin{figure*}
\centering
     \includegraphics[width=17cm,height=5cm]{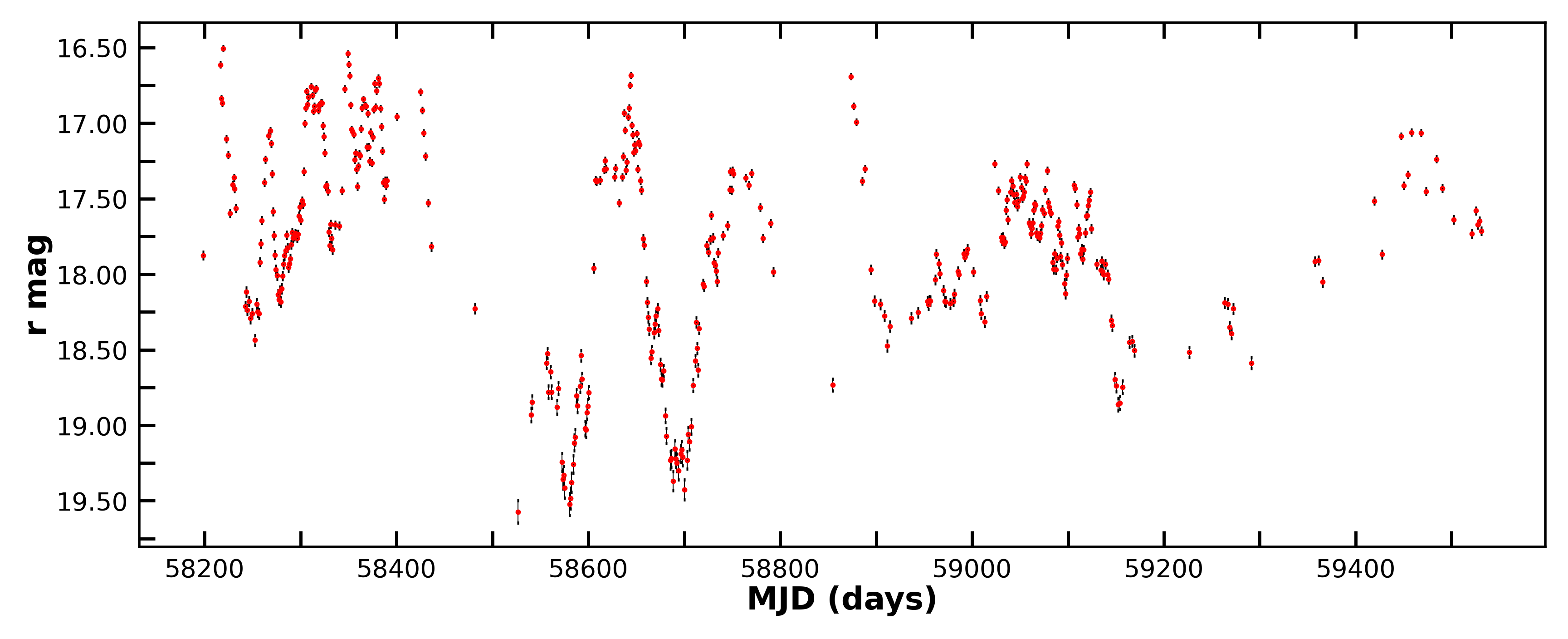}
    \includegraphics[width=17cm,height=10cm]{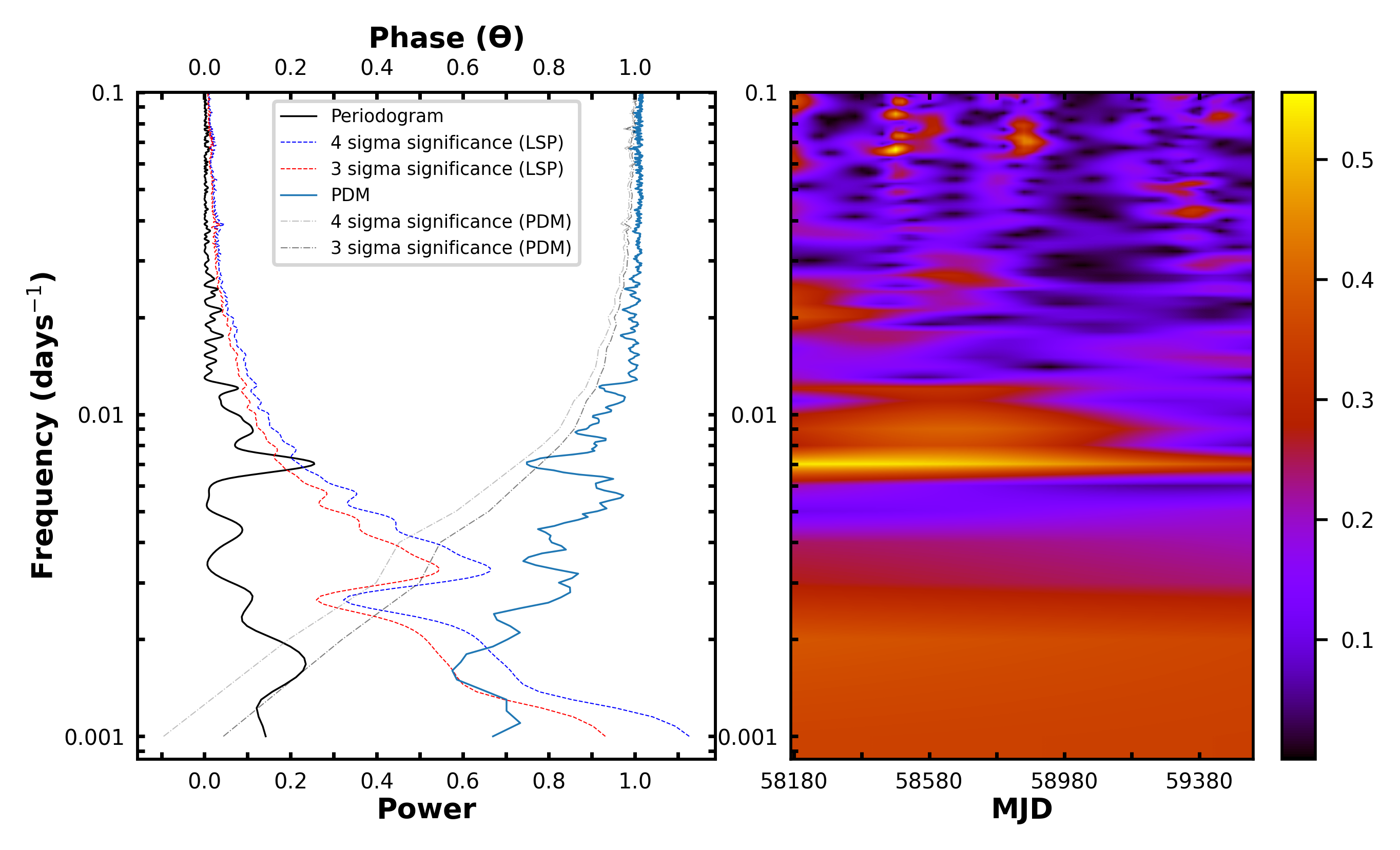}
    \caption{As in Figure~\ref{fig:J0929}, for the blazar J173927+495503. A low-frequency peak in the Lomb-Scargle periodogram at $\sim 4\sigma$ statistical significance was detected in the r-band light curve, suggesting the presence of a QPO at 0.0069 days$^{-1}$, and the phase in the PDM analysis also shows a dip at this frequency ($> 3\sigma$ significant). A power concentration is seen at that frequency in the WWZ plot over the entire span of observations but it weakens with time.}
    \label{fig:J1739}
\end{figure*}
\begin{figure*}
\centering
     \includegraphics[width=17cm,height=5cm]{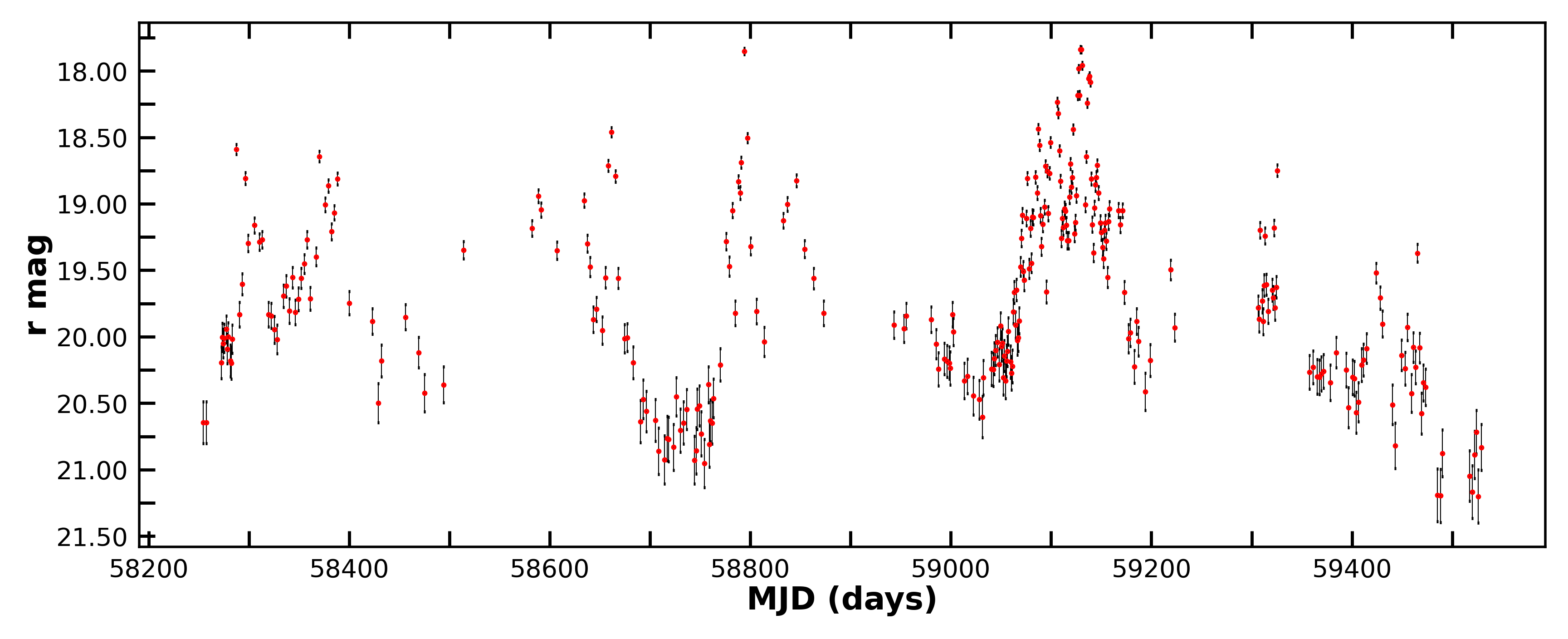}
    \includegraphics[width=17cm,height=10cm]{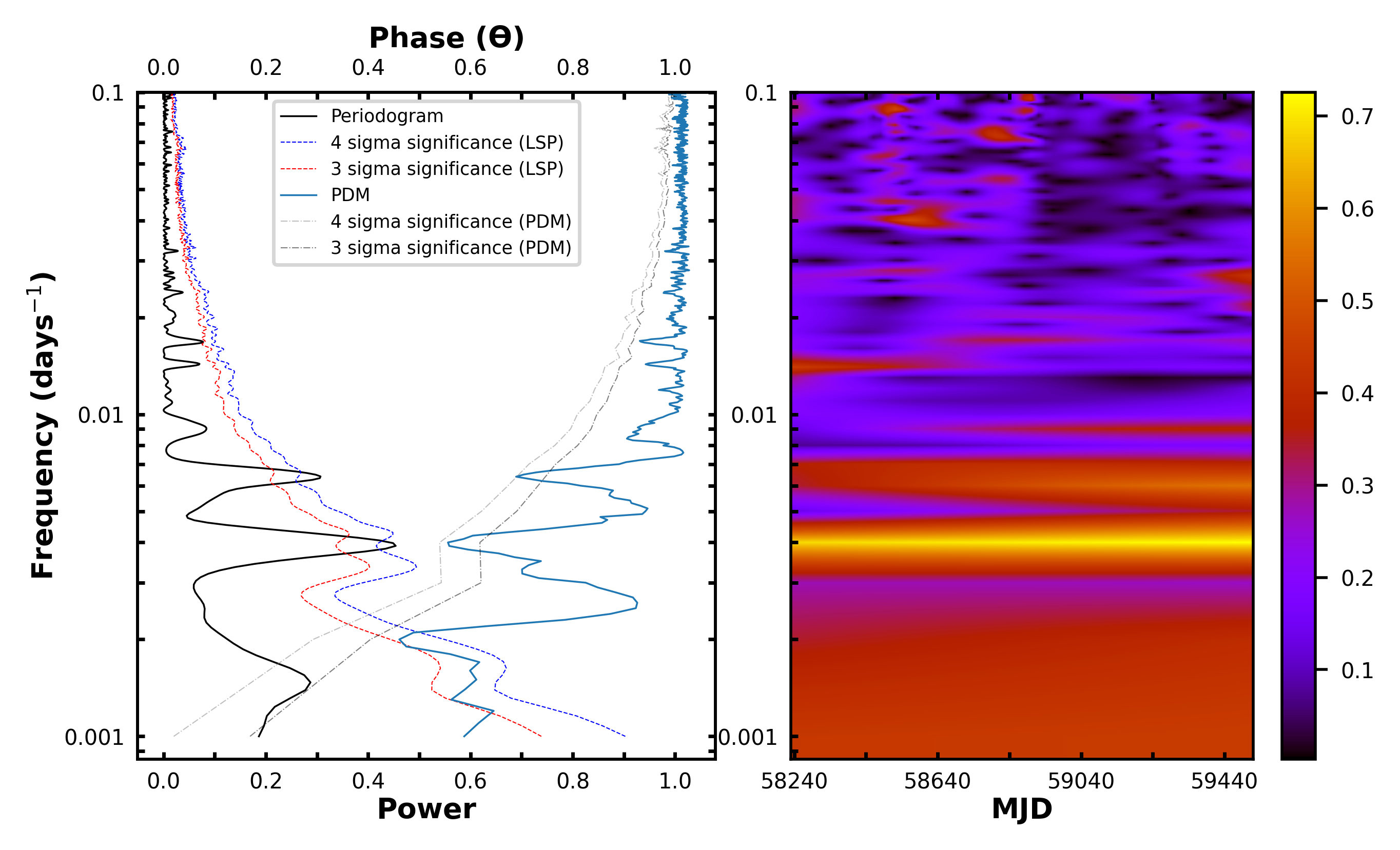}
    \caption{As in Figure~\ref{fig:J0929}, for the blazar J223812+274952. A low-frequency peak in the Lomb-Scargle periodogram at $\sim 4\sigma$ statistical significance detected in the r-band light curve, suggesting the presence of a QPO at 0.0065 days$^{-1}$. A dip in phase at that frequency ($>3\sigma$ significant) also is observed in the PDM analysis. A power concentration is seen at that frequency in the WWZ plot over the entire span of observations but it weakens with time. The other apparently significant peak seen at a frequency of $\sim$0.0039 is found to be a false signal, matching with the fake signal due to the data gap.}
    \label{fig:J2238}
\end{figure*}
We detect a strong signature of a QPO in a frequency range where these data are most useful in five blazars over their  $\sim$3.8 year long ZTF light curve data-sets. A  peak at relatively low-frequencies, ranging from $0.00509~ \text{days}^{-1}$ (i.e. 196 days) to $0.00604~ \text{days}^{-1}$ (i.e. 144 days) is detected in the Lomb-Scargle periodogram with $\gtrsim
 4\sigma$ significance in $r$-optical passband (see, Figures~\ref{fig:J0929}--\ref{fig:J2238}) in the case of 5 sources listed in Table~\ref{table:results_table_qpo}. 
 The QPO signature is also independently confirmed with the weighted wavelet transform method (see Figures~\ref{fig:J0929}--\ref{fig:J2238}), which makes them all strong cases of elusive persistent low-frequency QPOs. Further, we also checked for the QPO signatures in the g-band light curves of these 5 blazars. All the cases were found to show similar features in those power spectra with at least $\gtrsim3\sigma$ significance (see, Figures~\ref{fig:J0929_g}--\ref{fig:J2238}). A few other blazars showed comparably strong signals but were not further considered either because the peak was at too low a frequency or because the power was spread over a wide band of frequencies.  
 
 \subsection{Phase Dispersion Minimization Technique}
 This is another well-known technique for periodicity detection which is especially suitable for non-sinusoidal waveforms \citep{PDM1978}. For a time series of $N$ data points represented as ($x_i, t_i$) where $x_i$ is the flux state at some particular instant $t_i$, the flux variance is given by
 \begin{equation}
     \sigma^2 = \frac{\Sigma(x_i - \Bar{x})^2}{N - 1}, ~ (i = 1,...,N)
 \end{equation}
 where $\Bar{x}$ is the flux mean. The idea is then to sub-divide the light curve into $M$ number of phase bins according to a trial period ($\phi$) such that each segment contains $n_j$ number of points that are similar in phase. If the $k^{th}$ data point in the $j^{th}$ bin is denoted by $x_{kj}$, then the sample variance within each phase bin is denoted as
 \begin{equation}
     s_j^2 = \frac{\Sigma(x_{kj} - \Bar{x_j})^2}{n_j - 1} ~~ (j = 1,...,M).
 \end{equation}
 The overall variance including all the phase bins reads as
 \begin{equation}
     s^2 = \frac{\Sigma(n_j - 1)s_j^2}{\Sigma{n_j} - M} ~~ (j = 1,...,M).
 \end{equation}
The periodogram statistic, $\theta = \frac{s^2}{\sigma^2}$, defined as the ratio between the sample variance to the overall variance provides a measure of scatter of the sample variance around the mean of the time series. For a true period the scatter is expected to be small, so $\theta$ will be low. On the other hand, for a false period sample variance approaches true variance, and $\theta$, therefore, approaches unity. After plotting the test statistic $\theta$ with each trial period, one obtains the local minimum $\theta_{\text{min}}$ indicating a frequency corresponding to the least scatter about the mean. \par 
However, using extensive Monte Carlo simulations to test the detection feasibility of a true positive periodicity against the red noise background (which is the case for AGN), it was concluded that the PDM technique has to be used very cautiously and QPO components occurring beyond one-third of the light curve duration tend to be false positives \citep{pdm2021krishnan}. Keeping this in mind, we have used the PDM technique only as an auxiliary tool to verify the putative periodicities obtained from WWZ and LSP techniques. We reject any dip occurring at a frequency corresponding to $>1/3$rd of the light curve duration and also exclude the false positives arising out of the data gaps by matching with the template obtained from WWZ false positive. This leaves us with only the true positive signals, and we find the QPO frequencies from the PDM match well with those obtained from LSP and WWZ in the case of both the r- and g-band data. In our PDM analysis, we have utilized the publicly available \textsc{pyastronomy} PDM 
software\footnote{\url{https://pyastronomy.readthedocs.io/en/latest/pyTimingDoc/pyPDMDoc/pdm.html}}. In Figures~\ref{fig:J0929}--\ref{fig:J2238} where we observe the signature of a significant peak from the LSP and WWZ methods, we detect strong dips in PDM phase as well, which further strengthens the claims of QPO detections. We tested the same method for g-band data as well, and the periodicity obtained in LSP and WWZ are also well corroborated with PDM in these g-band light curves (see Figures~\ref{fig:J0929_g}--\ref{fig:J2238_g}). Following the same method as in LSP and WWZ, we have estimated the significance values for the PDM dips, and in r-band data, we obtain a $3\sigma$ detection significance in all the cases, and $> 4\sigma$ detection significance in the case of J092331+412527 and J101950+63200. In the g-band data as well, the detection significance is $> 3\sigma$ in  all  cases, barring J092915+501336, for which the significance is marginally below 3$\sigma$.\par 
 
 Below, we discuss possible physical scenarios for the emergence of such a QPO feature. \par 


\section{Discussion and Conclusions}

For blazars, the approaching jet is aligned close to the observer's line of sight  ($< 10^{\circ}$; \citealt{1995PASP..107..803U}), and the entire electromagnetic spectrum is usually fully dominated by the emission from the jet. The extreme relativistic boosting means that the jet's emission appears strongly amplified, often overwhelming all the thermal emission contributed by the AGN's accretion disc and the host galaxy. Therefore, it is quite likely that the jet emission is the primary driver behind the observed QPO signature instead of the putative accretion disc or the coronal region. If the jet is precessing or if it possesses an internal helical configuration \citep{camenzind1992lighthouse,villata1999helical,rieger2004geometrical,mohan2015kinematics}, quasi-periodic flux variations could easily arise from the temporal variations in the
Doppler boosting factor as perceived by the observer. In the context of the blazar PG 1553+113, \citet{Ackermann_2015} described the possibility of a $\sim$2 year QPO in the $\gamma$-ray and
other waveband  emissions in this way. There could be several possible triggers behind the jet axis modulation, one being the Lense–Thirring precession of the disc  \citep[e.g.][]{fragile2009general}.\par 
    
It has been shown that if the AGN is part of a binary SMBH system, that could also induce the jet precession effect \citep{begelman1980massive,valtonen2008massive}, although such mechanisms are most likely to produce periodicities exceeding 1 year \citep{2007ralc.conf..276R}. Several such candidates demonstrating $\gamma$-ray QPOs with
periods longer than this ballpark have recently been reported and have found plausible explanations in this framework \citep{Ackermann_2015,sandrinelli2016quasi,sandrinelli2016gamma,zhang2017revisiting,zhang2017possible}. However, for jets closely aligned along the line of sight having large Lorentz factors, the detected periods could be significantly shorter \citep{rieger2004geometrical}. \citet{zhou201834} has  applied this framework in order to explain the 34.5 days QPO apparently detected in the BL Lac PKS 2247–131. In our cases as well, if the jets are closely aligned along the line of sight, and the Lorentz factors are high, such a jet precession effect could produce QPOs in the range between 144 and 196 days that we have obtained. \par 

The appearance of fast quasi-periodicity
in the jet emission could also be explained in the framework of magnetic reconnection events in nearly equidistant magnetic islands inside the jet \citep{huang2013magnetic}. Such a physical configuration could periodically boost the emitted radiation, which would likely be manifested as a rapidly transient QPO. \citet{shukla2018short} successfully applied this paradigm to model the extremely fast variability
($\sim$5 min) in the FSRQ CTA 102 during its outburst in 2016. This
model does appear to be capable of producing the 3.6-days QPO seen in $\gamma$-rays for PKS 1510$-$089 as well \citep{roy2022transient}. Current-driven instabilities near a recollimation shock seem to provide a good explanation for the recent detection of a 13 hr QPO in both optical and $\gamma$-ray fluxes from BL Lac \citep{2022Natur.609..265J}. However, the relatively longer optical QPOs exceeding 100 days we have found in these blazars do not seem to fit either of these paradigms.    \par 

We can also consider other non-axisymmetric phenomena like the rotation of an accretion disc hotspot around the innermost stable circular orbit \citep[e.g.][]{zhang1990rotation}, such that the optical flux modulation comes from the circular motion of the hotspot \citep[e.g.][]{gupta2019mnras}.
However, the $\sim$200 days period as observed in our QPO analysis is too long for this mechanism. In addition, the SMBH mass we obtain from the relevant relation 
\begin{equation}
    \frac{M}{M_\odot} = \frac{3.23\times{10^4}P}{(r^{3/2}+a)(1+z)},
\end{equation}
where $P$ is the period in seconds, $r$ is the distance of the hotspot in units of $GM/c^2$, $a$ is the spin parameter, and $z$ is the redshift, turns out to be $\sim$10$^{10}~M_\odot$, which is on the very high end of plausible values. Therefore, such a `rotating hotspot' interpretation is unlikely to explain the observed QPO in our case.
\par

A geometrical origin instead could also be considered to be a plausible explanation. At radio wavelengths, Very Long Baseline Interferometry (VLBI) has
revealed that in some blazars, the parsec-scale cores appear to be
misaligned with the large, kiloparsec-scale structures of jets \citep[e.g.][]{sarkar2019long}. One natural explanation of such misalignment could be intrinsic helical structures of these inner jets. Such helical structures could be quite common in blazar sources \citep[e.g.][]{villata1999helical}. The origin of such helical structures could be connected to the hydrodynamic instability effects in magnetized jet \citep[e.g.][]{hardee1999dynamics}, 
as well as the interaction of the plasma blobs with the proximate medium \citep{godfrey2012periodic}. Propagation of a relativistic shock or plasma blob out along such a perturbed jet could induce significantly enhanced emission as the shock passes through a region of enhanced electron density. Owing to the extreme Doppler boosting effect, such a flux enhancement would be quite pronounced for an observer looking down the jet and such an effect could naturally be connected to the genesis of a QPO feature \citep{camenzind1992lighthouse}. Such an intersection of a relativistic shock with successive twists
of a non-axisymmetric jet structure scenario was invoked by \citet{rani2009nearly} in order to reasonably explain the claimed 17 days QPO in AO 0235+164. Depending upon the pitch angle, the Lorentz factor, and the viewing angle, such a mechanism is also compatible with the observed several-month periods of the QPOs observed in the present cases. \par 
For the simple one-zone leptonic model where the plasma blob containing elevated particle density and magnetic energy density is moving along a postulated helical trajectory along the extension of the jet, the viewing angle changes with time as
\begin{equation}
    \cos{\theta_{\text{obs}}(t)} = \sin{\phi}\sin{\psi}\cos{2\pi{t}/P_{\text{obs}}} + \cos{\phi}\cos{\psi}
\end{equation}
where $\psi$ is the angle of the jet relative to our line of sight and $\phi$ is the pitch angle of the blob, and $P_{\text{obs}}$ is the observed periodicity \citep{sobacchi2016model}. Since the viewing angle varies with time, the Doppler factor also becomes time variable as $\delta = 1/[\Gamma(1-\beta\cos{\theta(t)})]$, where $\Gamma = 1/\sqrt{1-\beta^2}$ is the bulk Lorentz factor. Given this scenario, the periodicity in the rest frame of the blob is given by
\begin{equation}
    P_{\text{rf}} = \frac{P_{\text{obs}}}{1 - \beta\cos{\psi}\cos{\phi}}.
\end{equation}
For typical values of $\phi = 2^{\circ}$, $\psi = 2^{\circ}$ and $\Gamma = 10$, the rest frame periodicity becomes $\sim$97 years for $P_{\text{obs}} \sim 150$ days. During this period, the blob traverses a distance $D = c\beta P_{\text{rf}}\cos{\phi}$, which amounts to $\sim 30$ parsec during one period. Since for the detection of statistically significant QPO in the present case we are considering 4-5 cycles at least, during the domain of observation the blob would travel $\sim$150 pc. In order to explain the variation of amplitude during the domain of observation, \citet{roy2022transient} considered that the angle relative to the line of sight could be time-dependent, arising out of a geometrically curved jet. Over a length scale of $150$pc, such a curved jet structure can form and thereby drive quasi-periodicity with the time-dependent amplitudes we have obtained in our samples. Such a conclusion has been drawn by \citet{roy2022transient} for the case of PKS 1510$-$089. 

\section{Summary} We report highly probable detections of low-frequency QPOs with periods ranging from  144 days to 196 days in the $\sim$3.8 year long ZTF r-band optical light curve in five blazar sources. 
The QPO is also apparently present in the g-band light curve and lasts for the entire data set at $\gtrsim$3$\sigma$. Such QPO signals are most likely to originate from the precession of high Lorentz factor jets, closely aligned to the observer's line of sight or the movement of a plasma blob along a helical jet structure. More such detections, made with surveys at a high cadence, and preferably at multiple bands, would increase our understanding of the underlying physical processes triggering the oscillatory features. \par

\section*{Acknowledgements}
The authors thank the referee for their comments and suggestions that have improved the quality of the manuscript. LCH was supported by the National Science Foundation of China (11721303, 11991052) and the National Key R\&D Program of China (2016YFA0400702). This work is based on observations obtained with the Samuel Oschin Telescope 48-inch and the 60-inch Telescope at the Palomar Observatory as part of the Zwicky Transient Facility project. ZTF is supported by the National Science Foundation under Grant No. AST-2034437 and a collaboration including Caltech, IPAC, the Weizmann Institute for Science, the Oskar Klein Center at Stockholm University, the University of Maryland, Deutsches Elektronen-Synchrotron and Humboldt University, the TANGO Consortium of Taiwan, the University of Wisconsin at Milwaukee, Trinity College Dublin, Lawrence Livermore National Laboratories, and IN2P3, France. Operations are conducted by COO, IPAC, and UW.

This research has made use of the NASA/IPAC Extragalactic Database (NED) which is operated by the Jet Propulsion Laboratory, California Institute of Technology, under contract with the National Aeronautics and Space Administration.

\section{Data Availability}
The data used in this study is publicly available in the ZTF DR10.


\bibliographystyle{mnras}
\bibliography{references} 

\appendix
\section{Results for the g-band analysis of the 5 sources}
\begin{figure*}
\centering
    \includegraphics[width=17cm,height=5cm]{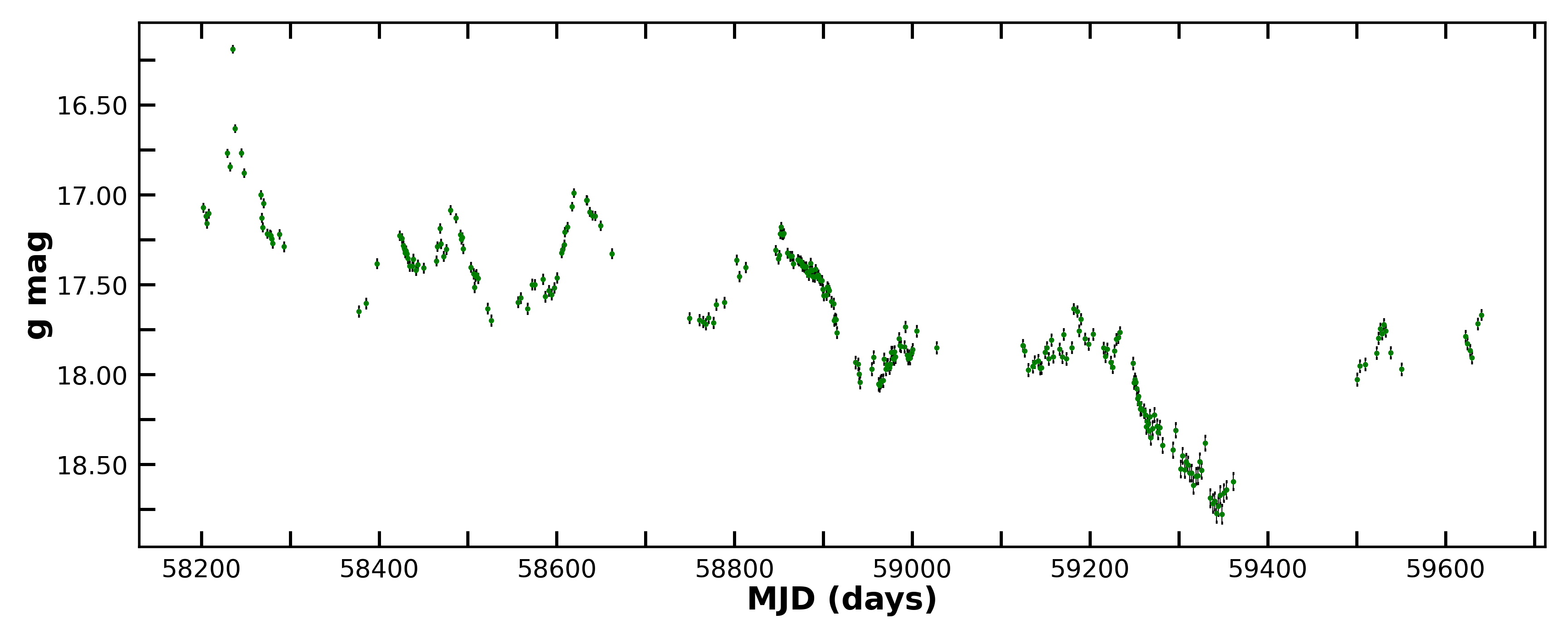}
    \includegraphics[width=17cm,height=10cm]{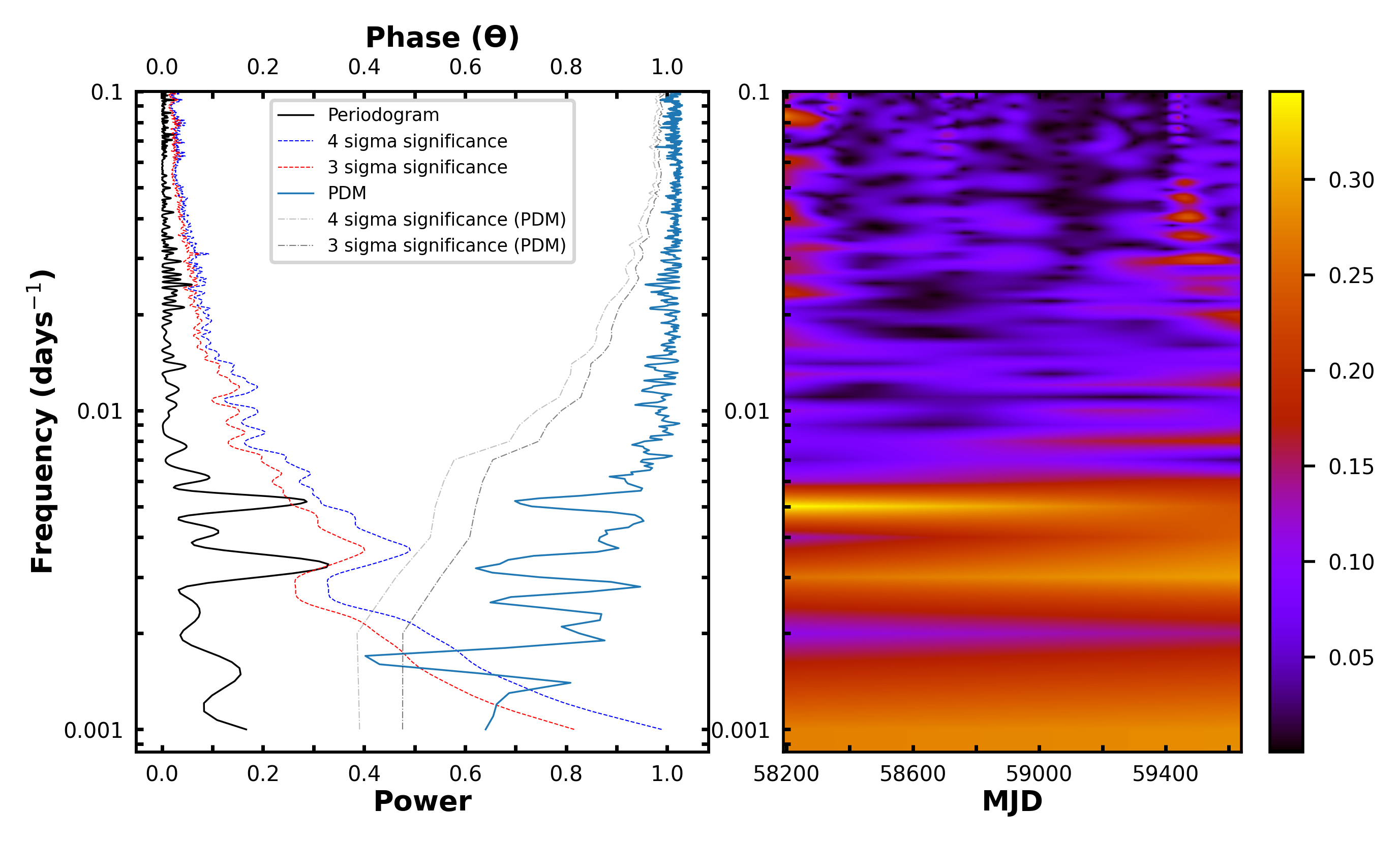}
    \caption{\textit{Upper Panel:} g-band ZTF light curve of blazar J092915+501336 over a time span of 3.8 years. \textit{Bottom left Panel:}  A low-frequency peak in the Lomb-Scargle periodogram at $> 3 \sigma$ statistical significance is detected in the g-band light curve, similar to the $> 4 \sigma$ peak in the r-band, suggesting the presence of a QPO at $\text{0.0051 d}^{-1}$. The PDM phase factor also shows a dip at that frequency, which is marginally below 3$\sigma$ significance level. \textit{Bottom right Panel:} A weighted wavelet Z-transform for the g-band optical light curve showing a strong power concentration over the entire observing time window at that slightly too low frequency.}
    \label{fig:J0929_g}
\end{figure*}

\begin{figure*}
\centering
    \includegraphics[width=17cm,height=5cm]{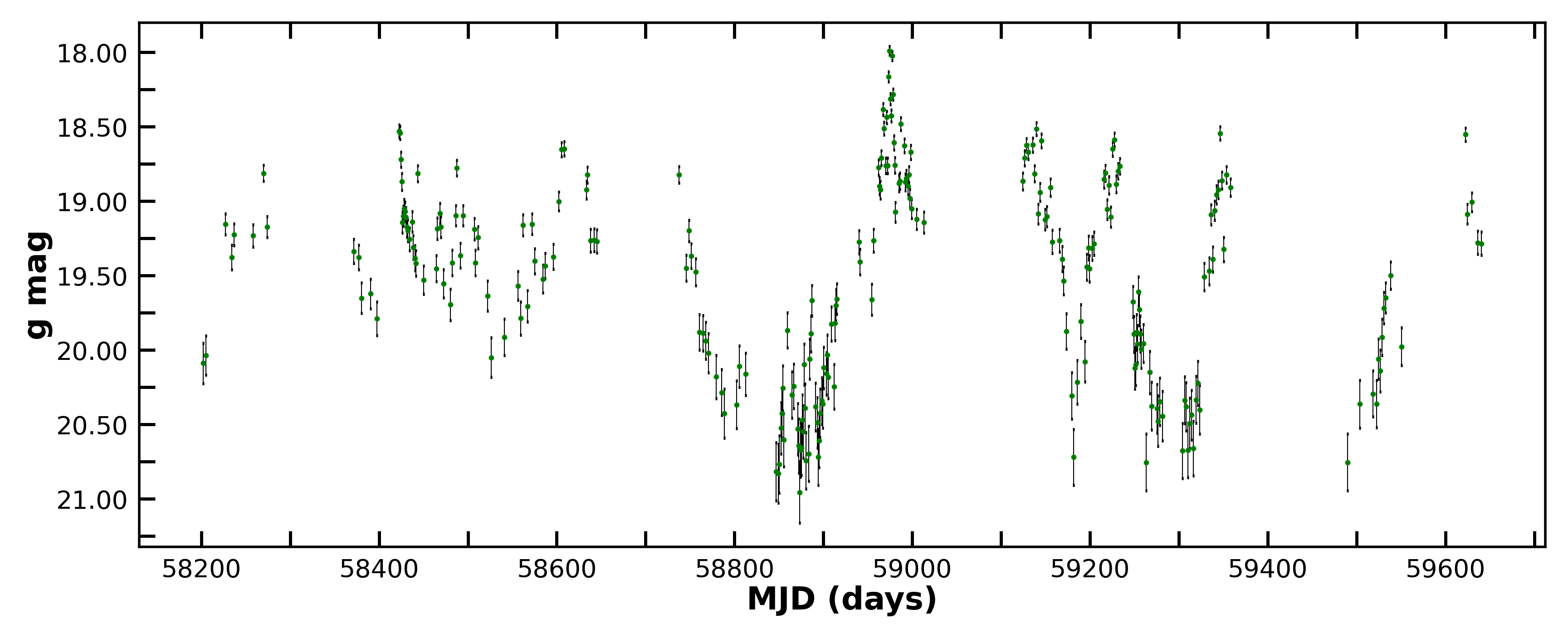}
    \includegraphics[width=17cm,height=10cm]{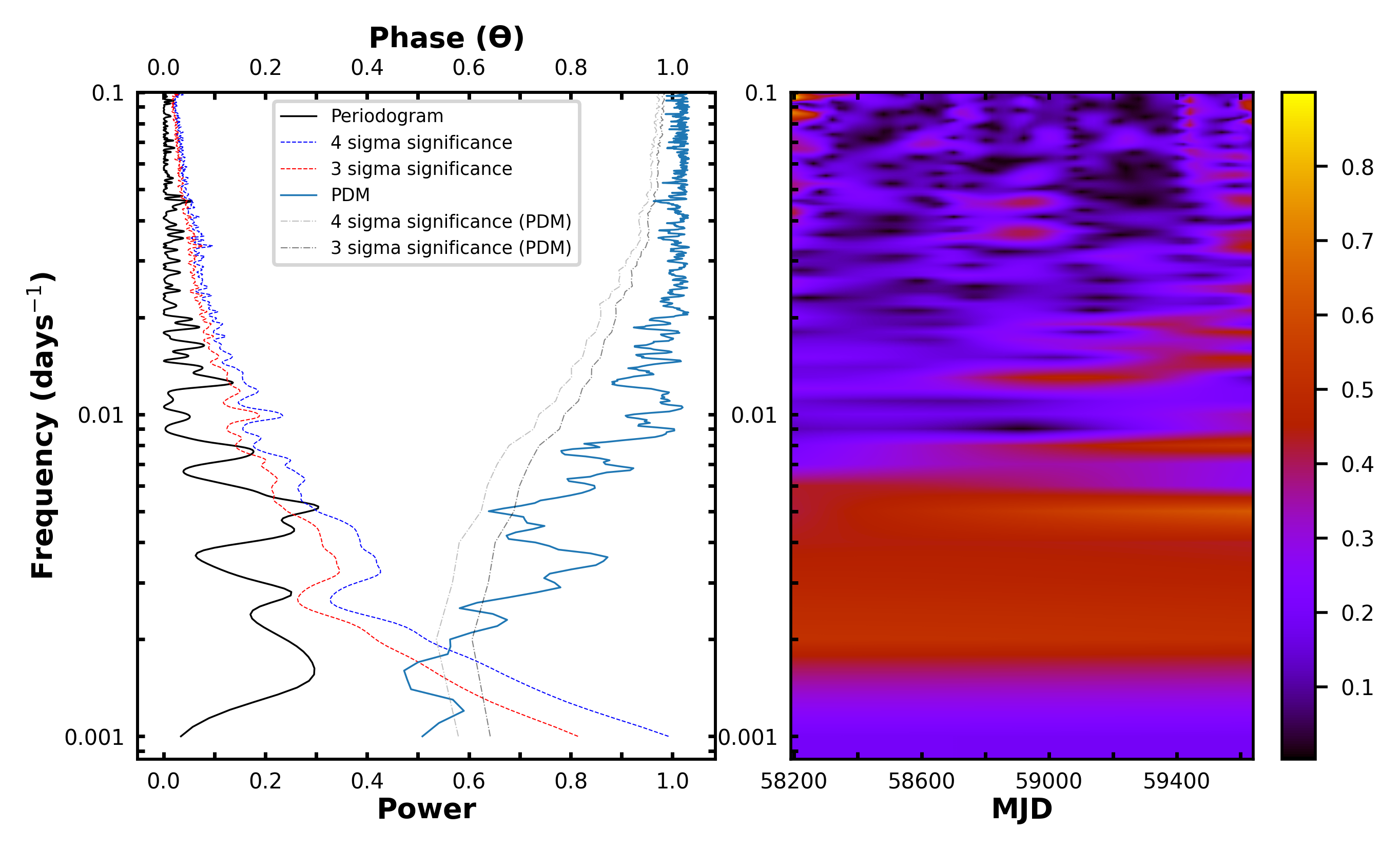}
    \caption{As in Figure~\ref{fig:J0929_g}, for the blazar J092331+412527. A broad low-frequency peak in the Lomb-Scargle periodogram at $> 4\sigma$ statistical significance detected in the g-band light curve, similar to the $> 4 \sigma$ peak in the r-band, suggesting the presence of a QPO around 0.0051 days$^{-1}$. A slump in the phase factor from the PDM analysis ($> 3\sigma$ significant) is also seen there. A  moderately strong but very broad power concentration is seen in the WWZ throughout the length of the observations. }
    \label{fig:J0923_g}
\end{figure*}

\begin{figure*}
\centering
    \includegraphics[width=17cm,height=5cm]{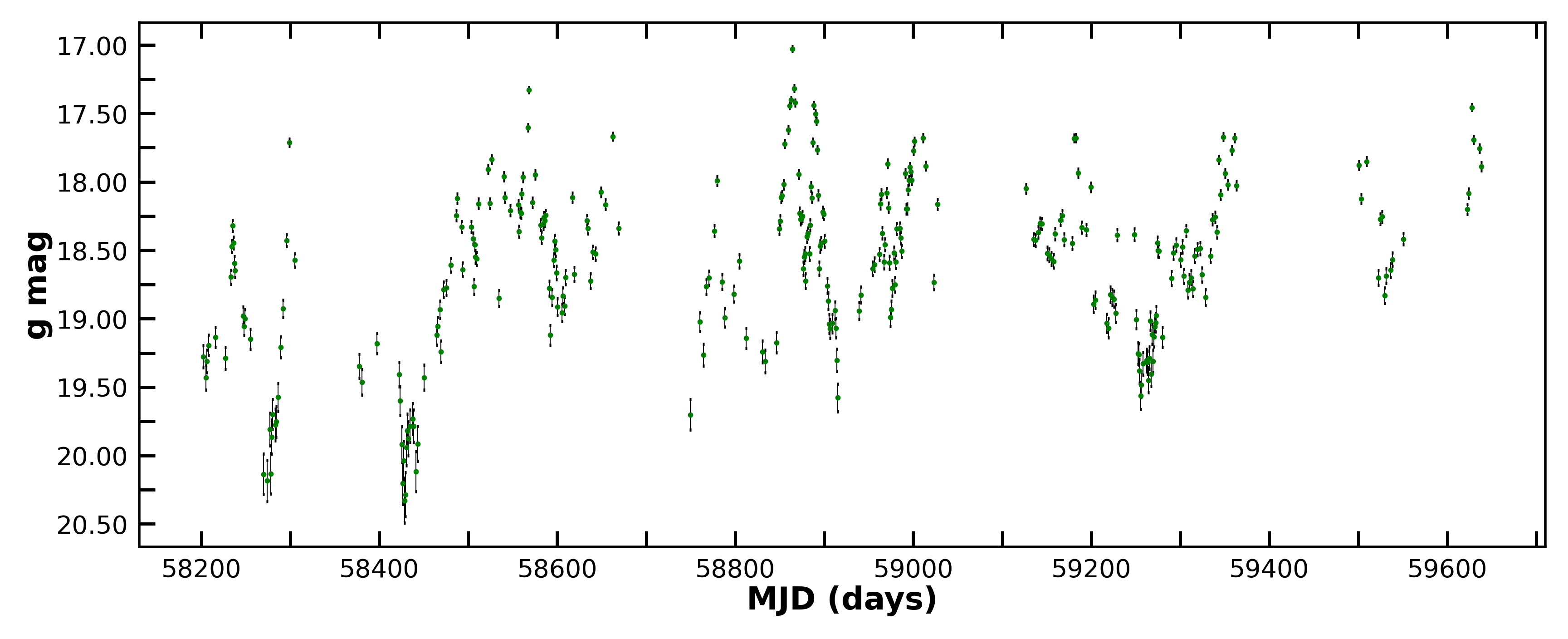}
    \includegraphics[width=17cm,height=10cm]{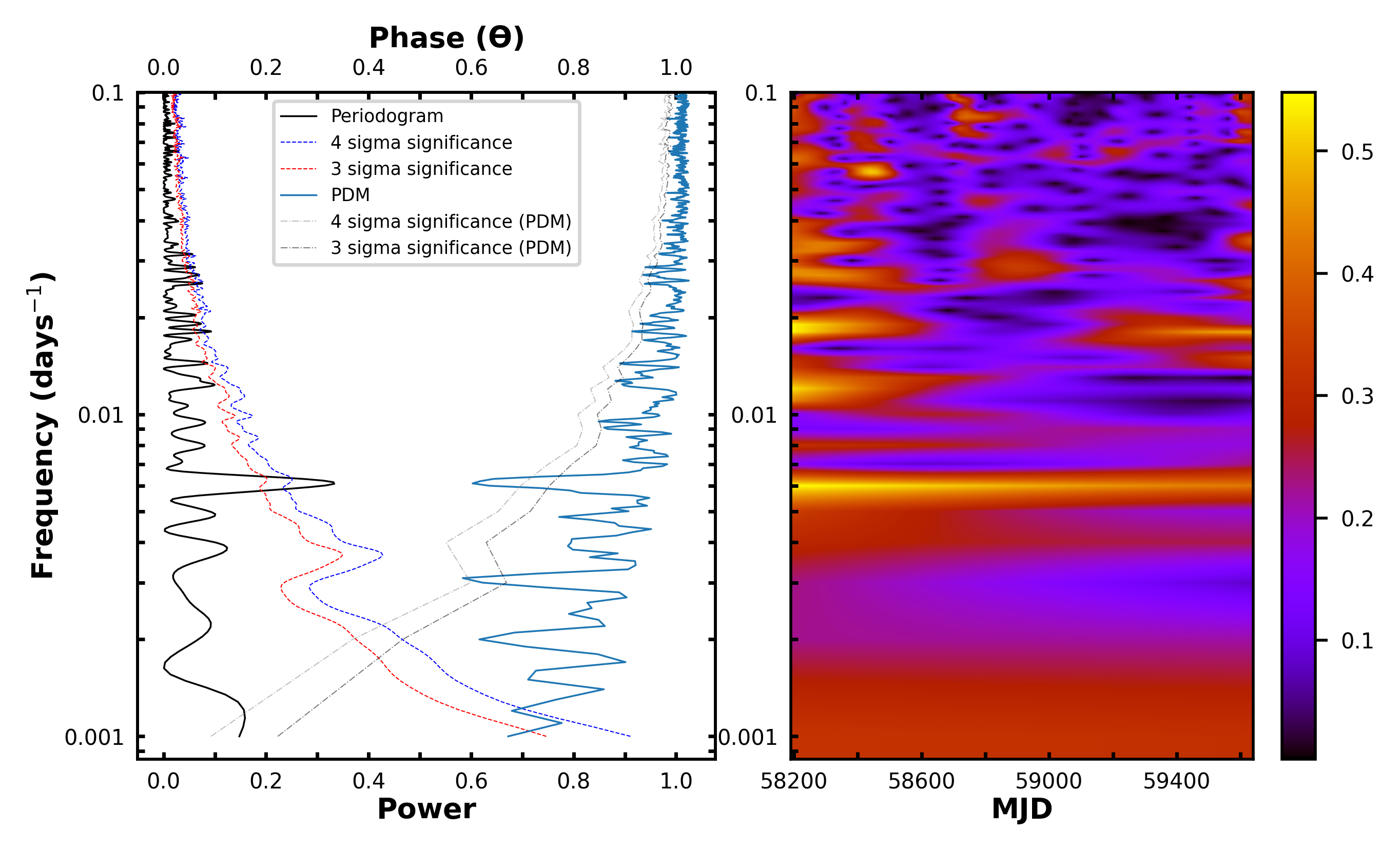}
    \caption{As in Figure~\ref{fig:J0929_g}, for the blazar J101950+632001. A low-frequency peak in the Lomb-Scargle periodogram at $> 4\sigma$ statistical significance detected in the g-band light curve, similar to the $> 4 \sigma$ peak in the r-band, suggesting the presence of a QPO around 0.0062 days$^{-1}$. A sharp drop in the phase factor to below 0.6 ($> 4\sigma$ significant) is observed in the PDM analysis. A strong and very narrow power concentration is seen in the WWZ throughout the length of the observations. }
    \label{fig:J1019_g}
\end{figure*}

\begin{figure*}
\centering
    \includegraphics[width=17cm,height=5cm]{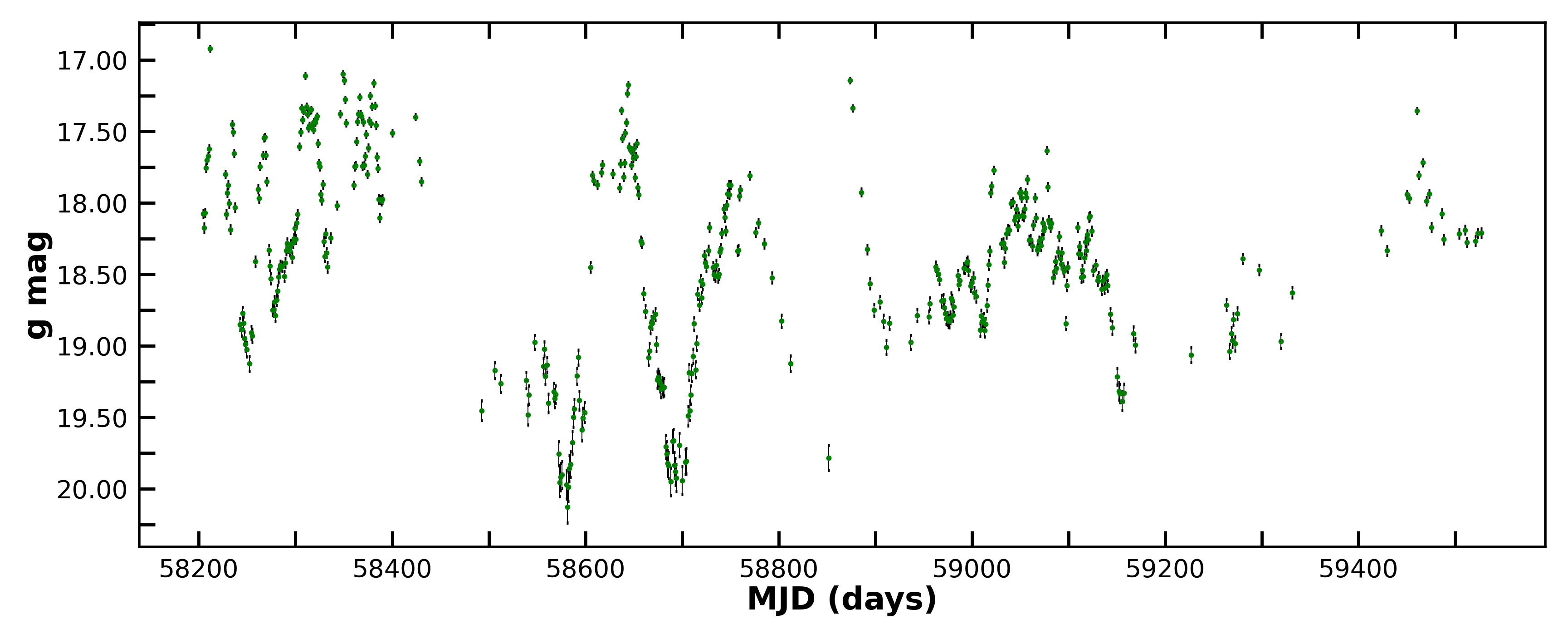}
    \includegraphics[width=17cm,height=10cm]{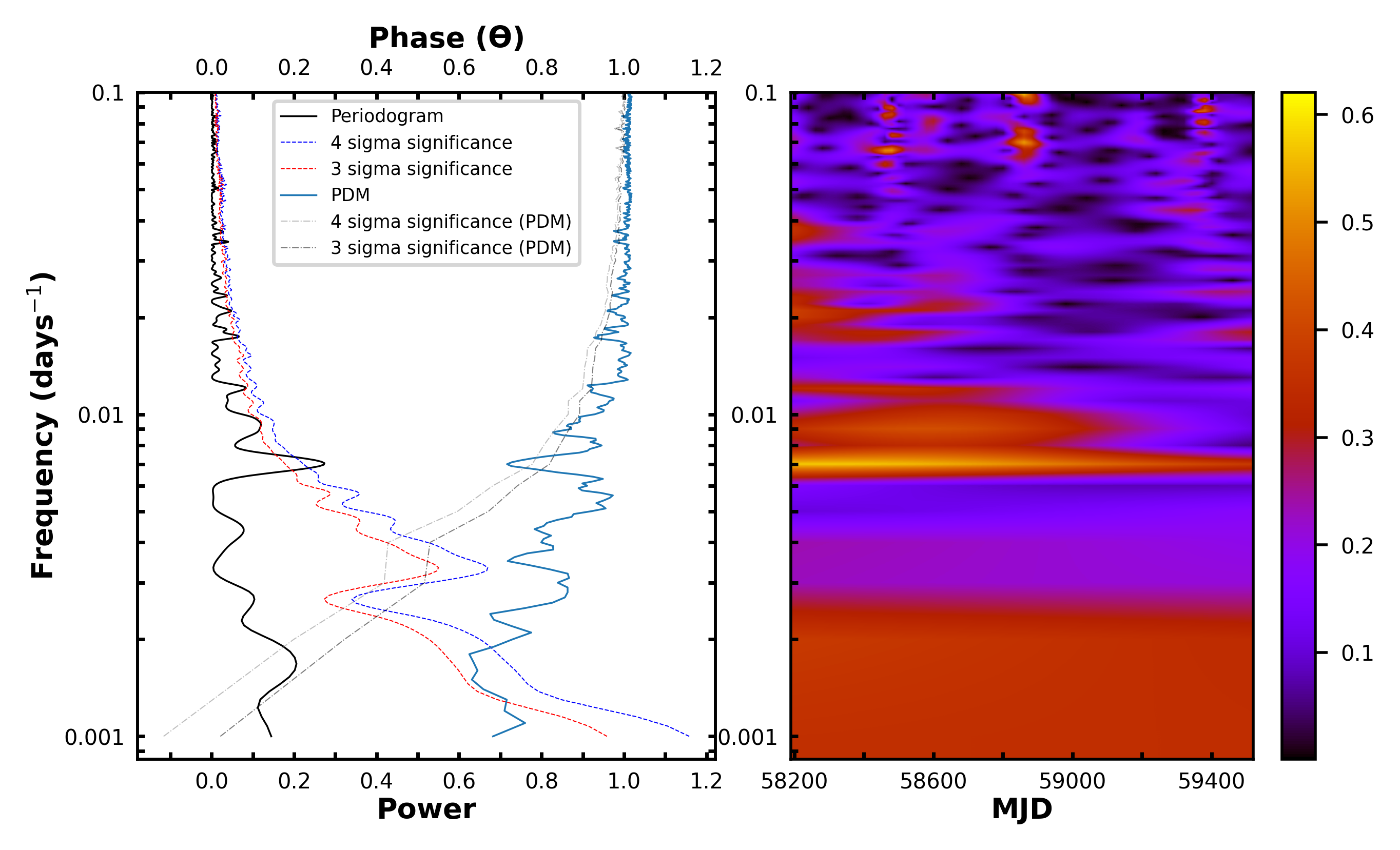}
    \caption{As in Figure~\ref{fig:J0929_g}, for the blazar J173927+495503. A low-frequency peak in the Lomb-Scargle periodogram at $> 4\sigma$ statistical significance detected in the g-band light curve, similar to the $> 4 \sigma$ peak in the r-band, suggesting the presence of a QPO around 0.0069 days$^{-1}$. A minimum in the phase factor is detected in the PDM analysis ($> 4\sigma$ significant) at that frequency. A strong and very narrow power concentration is seen in the WWZ throughout the length of the observations. }
    \label{fig:J1739_g}
\end{figure*}

\begin{figure*}

\centering
    \includegraphics[width=17cm,height=5cm]{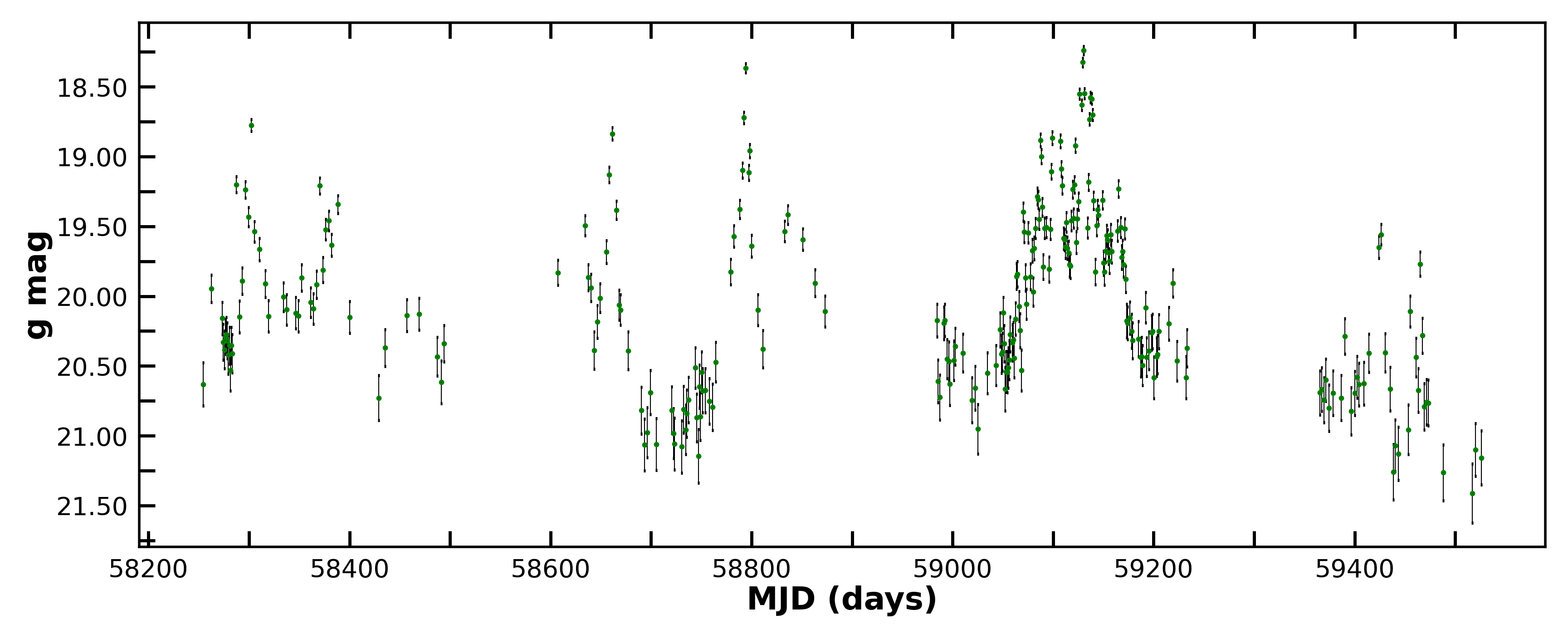}
    \includegraphics[width=17cm,height=10cm]{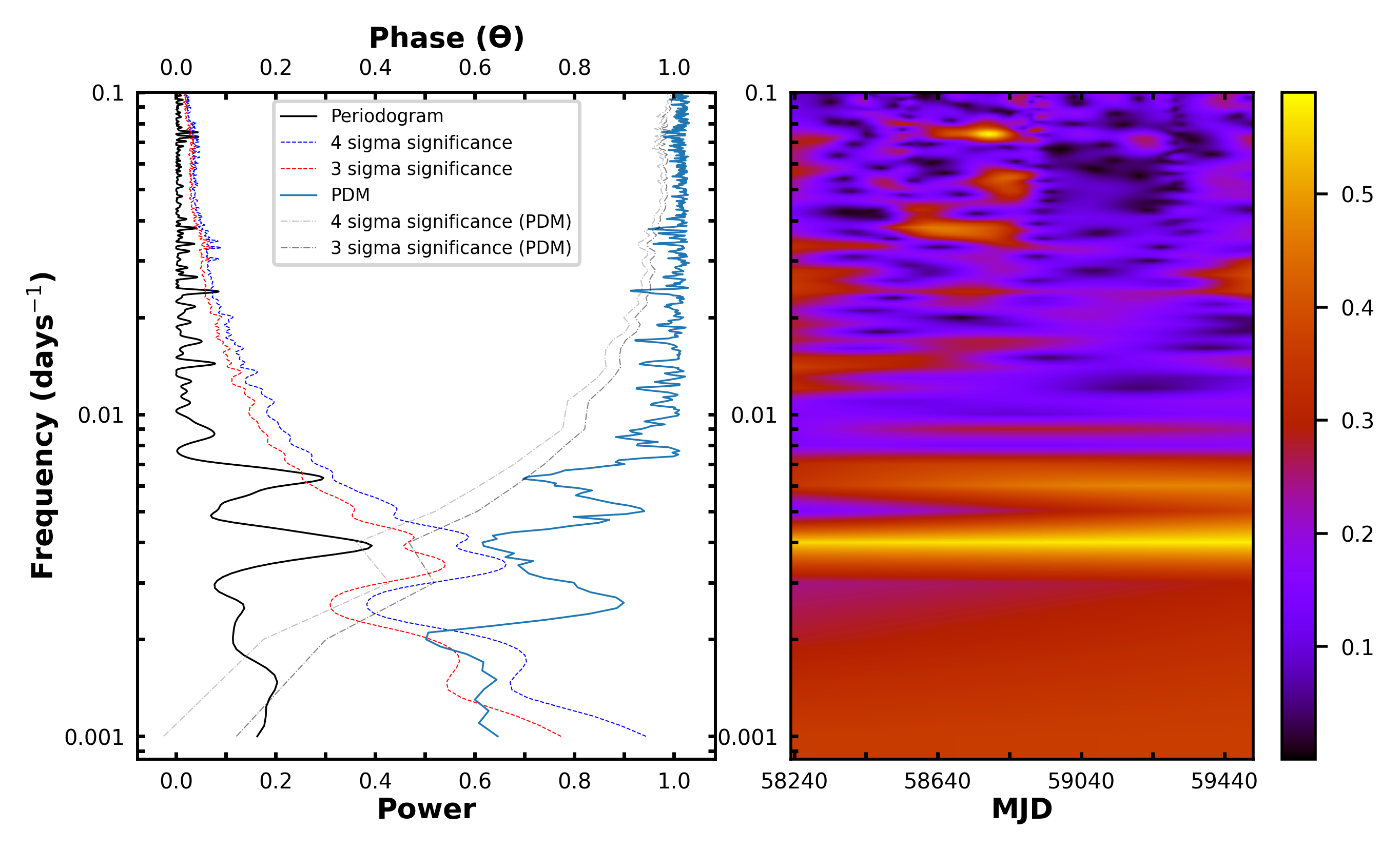}
    \caption{As in Figure~\ref{fig:J0929_g}, for the blazar J223812+274952. A low-frequency peak in the Lomb-Scargle periodogram at $> 3\sigma$ statistical significance detected in the g-band light curve, similar to the $> 4 \sigma$ peak in the r-band, suggesting the presence of a QPO around 0.0065 days$^{-1}$. At this frequency, a prominent dip in the phase factor is also observed from our PDM analysis ($> 3\sigma$ significant). A strong and very narrow power concentration is seen in the WWZ throughout the length of the observations. }
    \label{fig:J2238_g}
\end{figure*}

In this section, we present  information on the analysis of the g-band light curves of the 5 sources showing a prominent peak/dip in the WWZ, LSP and the PDM analyses. As mentioned earlier, we are considering only those cases to be legitimate ones for which a peaked signature is observed in the WWZ map (i) above a frequency of 0.005 $\text{days}^{-1}$ and (ii) concentrated within a narrow window of frequency and present throughout the observed duration. The analysis of the g-band light curve of these 5 sources showed similar signatures of periodicity (see, Figures~\ref{fig:J0929_g}--\ref{fig:J2238_g}), which further confirms the true nature of these QPOs.


\label{lastpage}
\end{document}